# Is the medical image segmentation problem solved? A survey of current developments and future directions

Guoping Xu[1], Jayaram K. Udupa[2], Jax Luo[3], Songlin Zhao[4], Yajun Yu[1], Scott B. Raymond[3], Xiaoxue Qian[1], Nian Wang[5], Hao Peng[1], Steve Jiang[1], Weiguo Lu[1], Lipeng Ning[6], Yogesh Rathi[6], Wei Liu[4], You Zhang[1#]

[1]The Medical Artificial Intelligence and Automation (MAIA) Laboratory, Department of Radiation Oncology, University of Texas Southwestern Medical Center, Dallas, TX 75390, USA

[2]Medical Image Processing Group (MIPG), Department of Radiology, University of Pennsylvania, Philadelphia, PA 19104, USA

[3]Neurological Institute, Cleveland Clinic, Cleveland, OH, 44195, USA

[4]Department of Radiation Oncology, Mayo Clinic, Phoenix, 85054, AZ, USA

[5]Advanced Imaging Research Center, University of Texas Southwestern Medical Center, Dallas, TX 75390, USA

[6]Brigham and Women's Hospital, Harvard Medical School, Boston, MA 02115, USA

[#]Email: You.Zhang@UTSouthwestern.edu

## Abstract

Medical image segmentation has advanced rapidly over the past two decades, driven largely by deep learning, enabling accurate and efficient segmentation of organs, tissues, cells, and pathologies across diverse modalities. This progress raises a fundamental question: to what extent have current models overcome persistent challenges, and what gaps remain? In this work, we provide an in-depth review of medical image segmentation, tracing its progress and key developments over the past decade. We discuss core principles—including multiscale analysis, attention mechanisms, and the integration of prior knowledge—across the encoder, bottleneck, skip connections and decoder of deep neural networks. Furthermore, we organize our discussion around seven key dimensions: (1) the shift from supervised to semi-/unsupervised learning, (2) the transition from organ to lesion segmentation, (3) advances in multi-modality and domain adaptation, (4) the role of foundation models and transfer learning, (5) the move from deterministic to probabilistic segmentation, (6) the progression from 2D to 3D and 4D segmentation, and (7) the trend from model-invocation to segmentation agents. Our goal is to provide a comprehensive perspective on the trajectory of deep learning-based medical image segmentation and to inspire further innovation. To support ongoing research, we will maintain a continually updated repository of relevant literature and open-source resources at https://github.com/apple1986/medicalSegReview.

## 1. Introduction

Medical image segmentation—the process of assigning a categorical label to each pixel or voxel in an image—plays an essential role in computer-aided diagnosis, disease detection and monitoring, treatment planning, and outcome assessment [1, 2]. Before the advent of deep learning, segmentation methods for both general and medical imaging primarily evolved along two paradigms. The first, purely image-based approaches, relied solely on information extracted from the input image to group pixels or voxels into meaningful regions. Representative techniques include level set [3, 4], active shape [5], graph cut [6], fuzzy connectedness [7], and watershed transforms [8, 9], among others. The second category, model-based approaches, incorporated prior knowledge in the form of statistical shape and appearance models [10], geometric and fuzzy models [11], or atlases-based frameworks [12]. These classical methods laid the theoretical and methodological foundations for modern segmentation research and inspired subsequent advancements. Nonetheless, their accuracy, computational efficiency, generalization ability, and robustness—particularly when applied across diverse imaging modalities and clinical scenarios—remained limited, leaving substantial room for improvement.

With the advent of deep learning [13, 14], advancements in image segmentation have accelerated significantly [15-20]. Deep neural network-based approaches have emerged as pivotal tools in the medical domain due to their ability to analyze images with high accuracy and efficiency. Both deep learning–based methods and traditional segmentation approaches facilitate precise quantitative analyses, delivering objective and standardized metrics that are essential for clinical trials. Among the deep learning contributions to medical image segmentation, U-Net [15] stands out as a seminal architecture, demonstrating wide applicability across imaging modalities such as MRI, CT, and ultrasound [21]. However, the limited receptive field of convolution-based neural networks has motivated the development of enhanced methods. These approaches often incorporate multi-scale feature extraction and fusion [20], channel [22] or spatial attention mechanisms [23], and skip connections [24] to address the limitations. For instance, DeepLab leverages atrous spatial pyramid pooling (ASPP) to extract multi-scale features without increasing the number of parameters [20]. Similarly, the High-Resolution Network (HRNet) [25] maintains high-to-low resolution convolution streams in parallel and facilitates effective information fusion across multiple resolutions.

Unlike convolution-based methods that rely on fixed convolution kernels to learn features, the self-attention mechanism computes explicit global context and long-range dependency relationships between all elements in a sequence, regardless of their distance. In the Transformer [16], a fully self-attention-based architecture was initially introduced for sequence-to-sequence modeling in natural language processing (NLP) tasks, demonstrating superior performance in capturing long-range relationships. Following this success, the Vision Transformer (ViT) [26] was proposed for vision tasks, achieving state-of-the-art results in image classification through pre-training on large-scale datasets. Subsequently, Transformer-based architectures have been explored for semantic segmentation, leading to models such as SETR [27], Swin Transformer [28], TransUNet [29], and LeViT-UNet [30]. Notably, a segmentation model, called Segment Anything Model (SAM), was introduced as a foundation model for generic object segmentation [31]. Trained on 11 million 2D natural images and over 1 billion segmentation masks, SAM demonstrated strong segmentation performance with prompts such as points, boxes, masks, or texts, marking a new milestone in segmentation capabilities. Owing to its exceptional generalizability, SAM has inspired numerous adaptations for medical image segmentation, including MedSAM [32], LeSAM [33], and MA-SAM[34]. Recently, SAM 2 was introduced for both images and videos, offering improved accuracy and achieving six times faster performance than its predecessor for image segmentation [35].

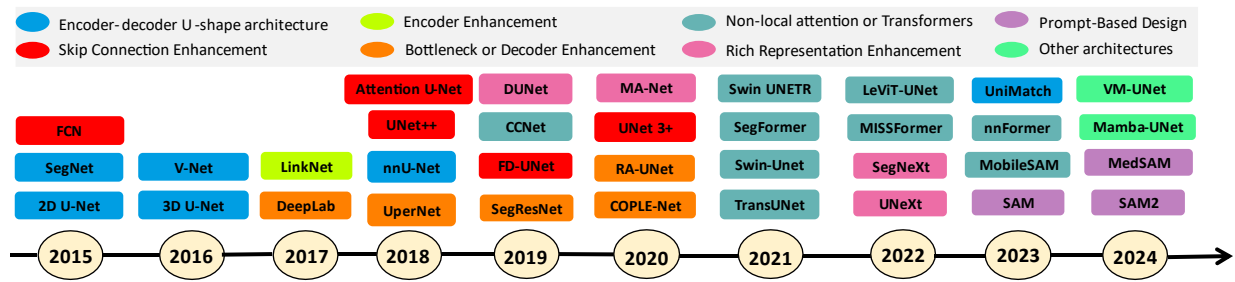

*Figure 1 Timeline of representative methods in image segmentation from 2015 to 2024. Inspired by [36], with acknowledgment to the original source.*

With the advancement of vision foundation segmentation models, a compelling question arises: ***to what extent have these models addressed the longstanding challenges of medical image segmentation, and what future directions remain to be explored in this domain?***

To address this question, we present a comprehensive survey tracing the evolution of medical image segmentation from fully supervised to semi-supervised deep neural network–based approaches. Our review encompasses a broad range of segmentation tasks, from organ delineation to tumor detection, across diverse medical imaging modalities and dimensional settings, including both 2D and 3D data. We focus on segmentation algorithms proposed between 2015 and 2024, with Figure 1 providing a timeline of representative methods over the past decade, illustrating the paradigm shift from convolution-based architectures to Transformer-based and hybrid convolution–Transformer models.

We begin by examining foundational design principles widely adopted in medical image segmentation, such as encoder–decoder architectures, multi-scale feature analysis, and attention mechanisms. Representative approaches are systematically summarized and analyzed, with an emphasis on the primary factors that have driven performance gains. For fully supervised segmentation, we use the vanilla U-Net as a baseline and review recent architectural innovations, highlighting advancements in the encoder, bottleneck, decoder, and skip connection designs. For semi-supervised segmentation, we discuss the theoretical underpinnings and practical implementations of key strategies, including consistency-based learning, pseudo-labeling, and the incorporation of prior knowledge. Their respective advantages, limitations, and applicability within the context of medical imaging are critically assessed.

Drawing on a decade of advances in deep learning for medical image segmentation, we organize the discussion around seven thematic dimensions that capture the field's current status and emerging directions: (1) challenges and opportunities in semi-supervised and unsupervised learning; (2) the shift from organ- to lesion-level segmentation; (3) progress from single- to multi-modality domain adaptation; (4) the influence of foundation models and transfer learning; (5) the evolution from deterministic to probabilistic segmentation; (6) the transition from 2D to 3D and 4D methodologies; and (7) the trend from model-invocation to segmentation agents. This framework provides a comprehensive synthesis and critically evaluates whether large-scale vision foundation models represent a transformative step toward overcoming persistent segmentation challenges, while outlining theoretical foundations and future research avenues for each dimension.

Existing surveys span a wide range of deep learning-based medical image segmentation methods, from U-Net variants [21, 36, 37] and supervised, weakly supervised [38], and semi-supervised approaches [39, 40] to Transformer-based [41] and SAM-based models [42-45]. Other reviews address specialized topics such as loss function design [46], imperfect dataset segmentation [47], generative adversarial networks [48], multi-modality fusion [49], and uncertainty analysis [50]. However, these reviews have not thoroughly explored the fundamental principles of image segmentation or comprehensively addressed both supervised and semi-supervised approaches to medical image segmentation. Recognizing this gap, we identified the need for a survey that bridges classical deep learning-based methodologies and the latest state-of-the-art techniques, including segmentation foundation models, while also highlighting future directions in medical image segmentation. Compared to previous studies, the primary contributions of this paper are summarized as follows:

- We present and summarize the foundational principles underlying deep learning-based medical image segmentation, such as encoder-decoder architectures, multi-scale analysis, and attention mechanisms.
- We analyze representative segmentation methods, aiming to identify the core innovations that drive performance improvements in both fully supervised and semi-supervised learning frameworks.
- We provide a comprehensive discussion on the current state of medical image segmentation and discuss future developments across seven critical dimensions.
- We compile seminal methods, source code, and datasets related to medical image segmentation, aiming to support and inspire further advancements in this field.

The remainder of this survey is structured as follows: Section 2 provides an overview of the prerequisite background and key concepts of image segmentation. Sections 3 and 4 review representative methods in fully supervised and semi-supervised segmentation, respectively. In Section 5, we discuss the challenges and future directions of medical image segmentation across seven critical dimensions. Finally, Section 6 concludes the survey.

## 2. Prerequisite

### 2.1 Encoder and decoder structure

The encoder-decoder architecture, also known as the autoencoder in unsupervised learning, has a rich history that spans multiple domains and has evolved into a cornerstone of deep learning fields such as computer vision [51], natural language processing [52], and medical image analysis [15]. This concept originated from the idea of mapping inputs to outputs through a compact intermediate representation. In this framework, the encoder learns a low-dimensional, representative feature set, while the decoder reconstructs the original input data from these compact features [53]. This approach has been widely explored for tasks such as dimensionality reduction and latent space modeling (see Figure 2).

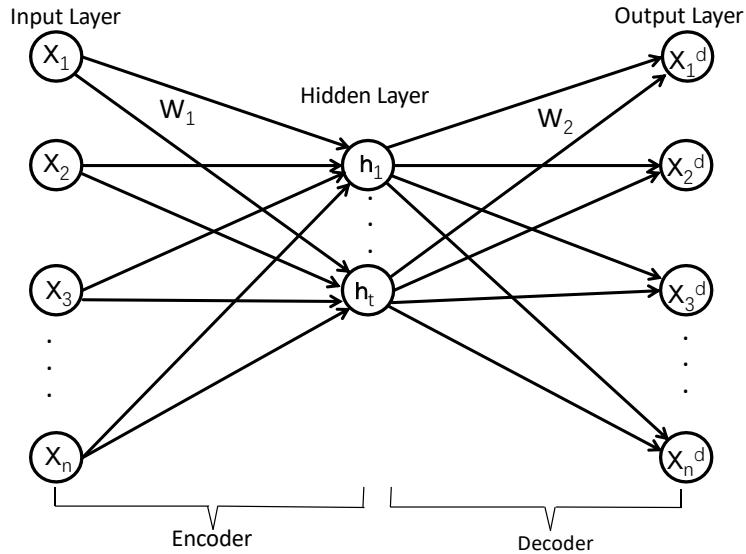

*Figure 2 The structure of encoder-decoder.*

Conventional encoder-decoder architectures initially utilized fully connected layers (multi-layer perceptrons) for processing one-dimensional data, as demonstrated in autoencoder [54], denoising autoencoder [55], and sparse autoencoder [56]. However, these architectures suffered from a significant limitation: converting 2D images into one-dimensional vectors for input led to the loss of spatial structure information during latent feature extraction. The advent of convolutional neural networks (CNNs) addressed this limitation with the introduction of convolutional autoencoders, which employed convolution and pooling layers in both the encoder and decoder [57, 58]. Concurrently, restricted Boltzmann machines (RBMs) [59] emerged as an effective method for initializing the weights of deep autoencoder networks, facilitating the learning of low-dimensional representative features [54, 60, 61] and accelerating the broader development of deep neural networks. Building on these advancements, encoder-decoder architectures found widespread application in image segmentation. The success of fully convolutional networks (FCNs) [19] paved the way for specialized encoder-decoder designs such as SegNet [52], LinkNet [62], and U-Net [15], which have become foundational models in medical imaging and related fields.

## 2.2 Multi-scale pyramidal architecture

The concept of multi-scale (or multiresolution) analysis is a fundamental approach in image processing and computer vision, enabling the examination of data at varying levels of resolution, granularity, or spatial/temporal scales. This technique facilitates the identification of features and patterns that might otherwise be overlooked when data is analyzed at a single scale. For example, a multiresolution signal decomposition framework based on wavelet representation was introduced in [63], and it has since been applied to a variety of tasks, including image compression [64], multi-scale texture mapping [65], and multi-scale object detection [66].

Traditionally, image pyramids [67] have been integral to conventional handcrafted feature-based methods for tasks such as feature extraction [68], denoising [69], and image registration [70]. With the rise of deep neural networks, the image pyramid or feature pyramid architecture has emerged as a cornerstone in advanced computer vision tasks, such as object detection [71], image restoration [72], and image segmentation [73, 74] (see Figure 3 for examples of multi-scale image and feature pyramid structures in deep learning). These developments underscore the importance of multi-scale analysis in enhancing both the effectiveness and robustness of traditional and deep learning-based methodologies. The following sections will provide a detailed discussion on multi-scale feature utilization in image segmentation.

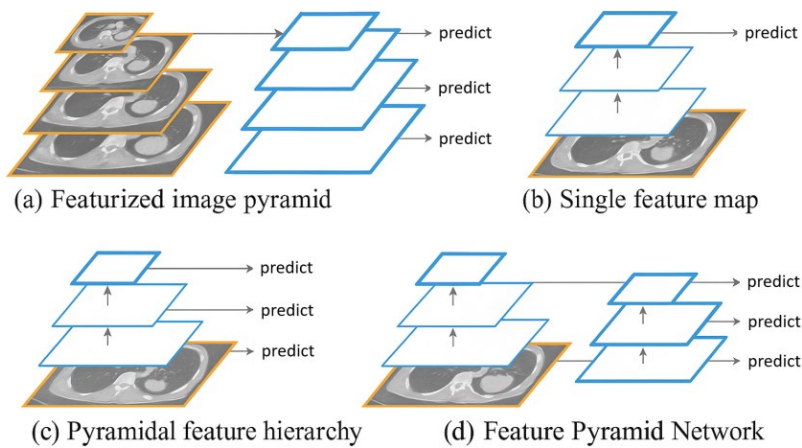

*Figure 3 Illustration of an image pyramid for multi-scale image or feature extraction in deep learning-based segmentation frameworks.*

A different form of multi-scale (more appropriately called morphological-scale or variable-scale) analysis concepts has also been developed in computer vision, which allows shift-variant operations on images. In the ball-scale method [75], for example, a ball-scale image is associated with every given image; the ball scale value at any pixel represents the radius of the largest ball (circular region) within which image intensity is considered to be uniform in the given image within an allowed limit to account for local noise. In tensor scale methods [76], the largest elliptical regions are considered, and in generalized-scale methods [77], the region is allowed to have any shape. These methods can be readily generalized to 3D imagery and textural properties. This knowledge of local scale was shown effective in a variety of image operations including filtering [75, 78], segmentation [79], registration [80], and image intensity standardization [81]. The utility of these variable scale methods in deep networks is yet to be fully exploited [82].

## 2.3 Attention mechanism

The attention mechanism is a cornerstone of modern deep learning, playing a pivotal role in both natural language processing [83, 84] and computer vision [26, 28]. It enables models to selectively focus on the most relevant parts of input data while disregarding less significant elements, effectively emulating the human ability to concentrate on critical information [85, 86]. In recent years, numerous attention mechanisms have been proposed to enhance the performance and versatility of deep neural networks. These mechanisms are broadly categorized into three types [87]: channel attention [22], naïve spatial attention [23], and self-attention [84]. Figure 4 presents a comparative overview of these mechanisms, illustrating the channel attention as implemented in the Squeeze-and-Excitation (SE) block, the naïve spatial attention in the Convolutional Block Attention Module, and the self-attention operation in Transformer. Channel attention generates a one-dimensional vector that assigns weights to individual feature maps, selectively emphasizing the most important channels. Spatial attention, on the other hand, produces a two-dimensional weight map and applies it uniformly across all feature maps to highlight spatially significant regions. Lastly, self-attention computes pairwise similarities between all entities across every feature map by multiplying the Query and Key matrices. This operation enables the model to capture long-range dependencies and contextual relationships, making self-attention particularly effective for tasks requiring a deep understanding of contextual relationships. In this survey, we will provide a detailed discussion of the attention mechanism in medical image segmentation.

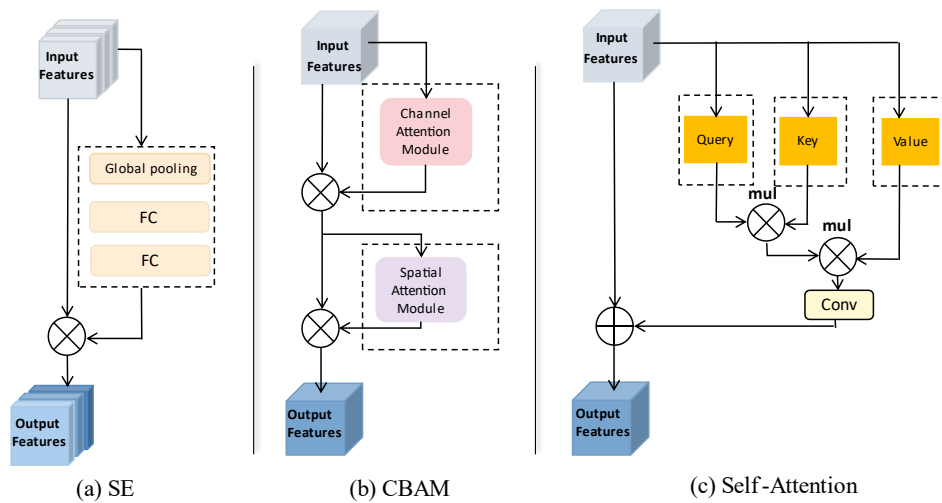

*Figure 4 Schematic comparison of various attention mechanisms: The Squeeze-and-Excitation block for channel attention (SE), the Convolutional Block Attention Module (CBAM), where channel and spatial attentions are connected in series, and self-attention, as first proposed in the Transformer model.*

## 2.4 Fully supervised, semi-/weakly-supervised, and unsupervised image segmentation

Image segmentation can be broadly categorized into three approaches based on the extent of annotations used: fully supervised learning, semi- and weakly-supervised learning, and unsupervised learning. Fully supervised segmentation typically leverages precise, pixel-level annotations for each target object, which in many studies has been shown to yield higher accuracy compared to weakly or semi-supervised approaches [19]. However, its performance can vary depending on factors such as annotation quality, dataset size, and the complexity of the target structures. Moreover, obtaining these annotations is time-intensive and expensive, particularly in medical imaging, where domain expertise is required.

To address these challenges, semi-supervised semantic segmentation leverages a small set of labeled images along with a larger set of unlabeled images for training. This paradigm has shown promise in accelerating the annotation process by reducing the volume of data requiring full manual labeling [88]. To further reduce the annotation burden, weakly-supervised methods have been developed, which utilize annotations that are less detailed or precise compared to full pixel-level labels. Examples of such annotations include image-level labels, bounding boxes, or scribbles. These weaker annotations offer a more efficient labeling process while still enabling models to learn meaningful segmentations [89]. In contrast, unsupervised semantic segmentation eliminates the need for labeled data entirely. This method focuses on partitioning an image into semantically meaningful regions or categories based solely on inherent visual features such as texture, color, and shape. Unlike supervised or weakly-supervised approaches, unsupervised segmentation discovers and groups patterns or objects in the image without any explicit guidance from human annotations [90]. Figure 5 illustrates examples of fully supervised, semi-supervised, weakly-supervised, and unsupervised segmentation applied to a CT image, highlighting the differences in annotation requirements and outcomes across these approaches.

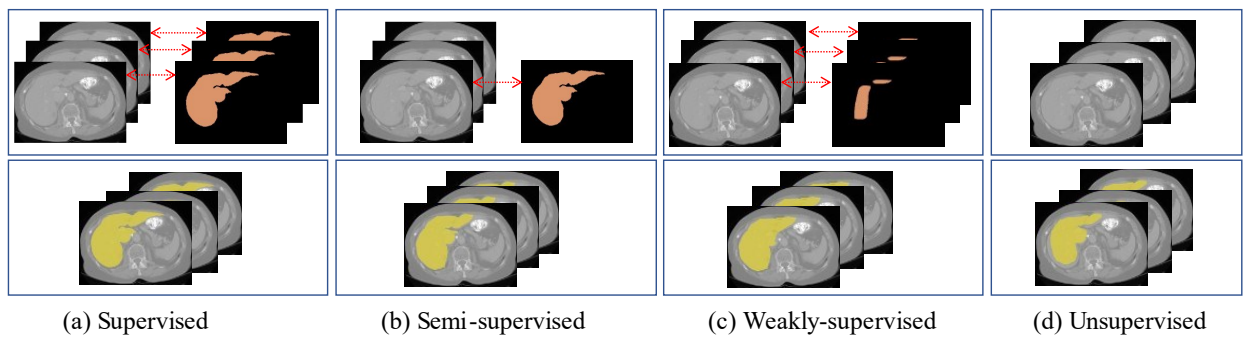

| (a) Supervised | (b) Semi-supervised | (c) Weakly-supervised | (d) Unsupervised |

*Figure 5 An illustration of supervised, semi-supervised, weakly supervised, and unsupervised learning for liver segmentation on CT images. The first row highlights differences in annotation quantity and quality, with orange labels indicating the available ground truth. From supervised to semi-supervised learning, the number of ground truth annotations decreases; from supervised to weakly supervised learning, annotation quality declines. The second row presents segmentation predictions (yellow) from*

## 3. Fully supervised segmentation

Fully Convolutional Networks (FCNs) were the pioneering deep learning models designed for semantic segmentation [19]. They use fully convolutional layers to handle inputs of arbitrary size and predict pixel-wise labels in an end-to-end manner. FCNs also introduce skip connections, which combine semantic information from deep layers with detailed local information from shallow layers. This integration is intended to enhance the segmentation performance by balancing coarse contextual and fine-grained details (See Figure 6).

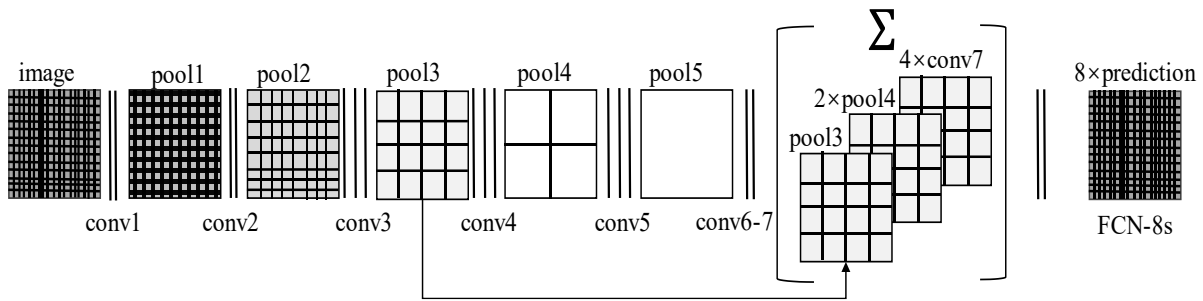

*Figure 6 The architecture of FCN-8s. It employs fully convolutional layers in an end-to-end manner. Additionally, it combines three feature maps from the outputs of pool3, pool4, and conv7, following interpolation, in an attempt to enhance precision.*

A modified and extended version of the Fully Convolutional Network (FCN), the U-Net [15] , was proposed to enhance segmentation performance. Compared to the FCN, the U-Net architecture comprises four key components: an encoder, a bottleneck, a decoder, and skip connections (see Figure 7). The *encoder* extracts semantic and contextual features, while the *bottleneck* learns a compressed representation of the input. The *decoder* progressively upsamples feature maps to produce finer segmentation results. *Skip connections* link the encoder and decoder, transferring localization information to aid the decoder in identifying and segmenting objects. By passing feature maps from the encoder to corresponding levels of the decoder, *skip connections* facilitate the integration of local and semantic information, improving overall segmentation performance [36]. Due to its simplicity and effectiveness, U-Net has become the de facto standard for biomedical image segmentation.

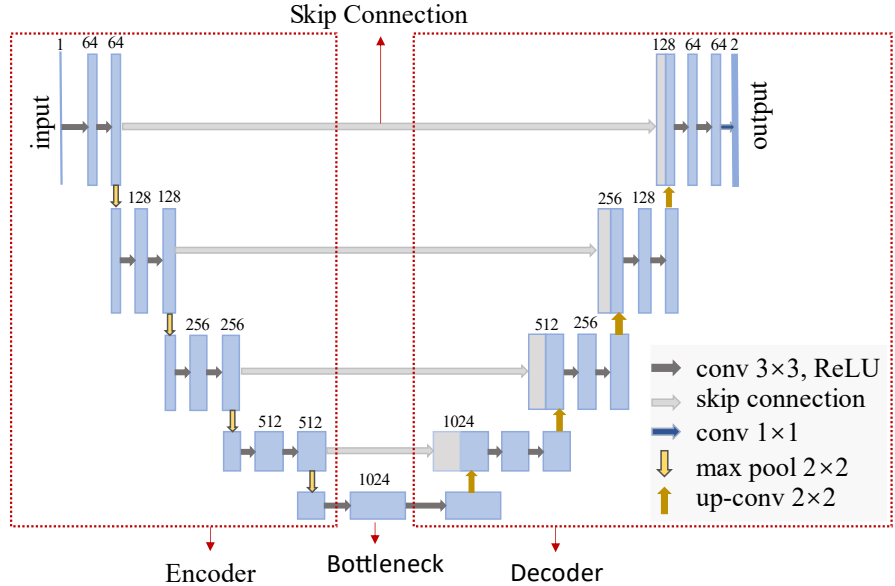

*Figure 7 Illustration of the U-Net architecture. The network comprises four main components: an encoder, a decoder, a bottleneck, and skip connections. Each blue box represents a multi-channel feature map, with the numbers above indicating the number of channels. Gray boxes denote feature maps copied from the encoder. Arrows in different colors represent various operations, as explained in the legend at the bottom-right.*

FCNs and U-Net have significantly advanced semantic segmentation through fully supervised learning, but they also exhibit several limitations across the encoder, bottleneck, decoder, and skip connections, as summarized in Figure 8. In the following sections, we will explore advancements aimed at addressing these limitations to achieve better segmentation performance.

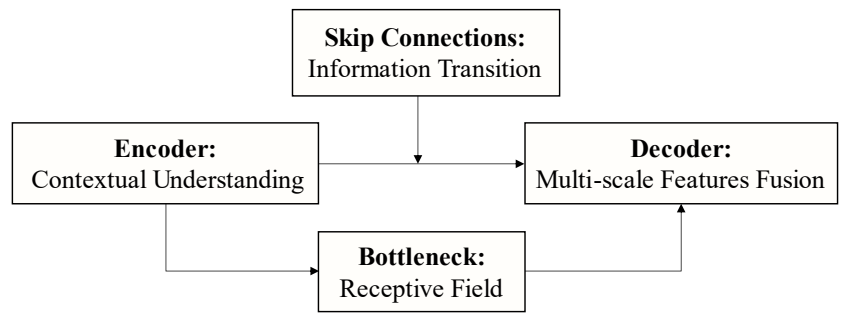

*Figure 8 Schematic illustration of the limitations of FCNs and U-Net, which include primarily limited contextual understanding in the encoder, a fixed receptive field in the bottleneck, challenges in multi-scale feature fusion within the decoder, and inefficient spatial information transition via skip connections.*

## 3.1 Contextual understanding in the encoder

Fully Convolutional Networks and U-Net rely on down-sampling operations (e.g., via pooling) to extract multi-scale features, which can lead to a loss of fine-grained spatial information. This limitation poses challenges in accurately delineating the boundaries of small or spatially sparse objects [91]. To address this issue, numerous works have proposed modifications to the vanilla FCN or U-Net encoder to extract richer and more fine-grained semantic representations [36]. These enhancements primarily focus on *integrating residual blocks or dense connections, variant convolution operations, and attention mechanisms.*

(1) *Residual blocks or dense connections in the encoder*

ResNet, originally proposed for image classification and object detection tasks [24], introduced a significant innovation compared to earlier networks like AlexNet [14] and VGG [92]. Instead of learning features directly, ResNet reformulates convolutional layers to learn residual features, which simplifies optimization and allows deeper networks to achieve higher accuracy [93]. Figure 9 (b) illustrates the key differences between ResNet and other networks, highlighting its innovative use of skip connection for residual feature learning.

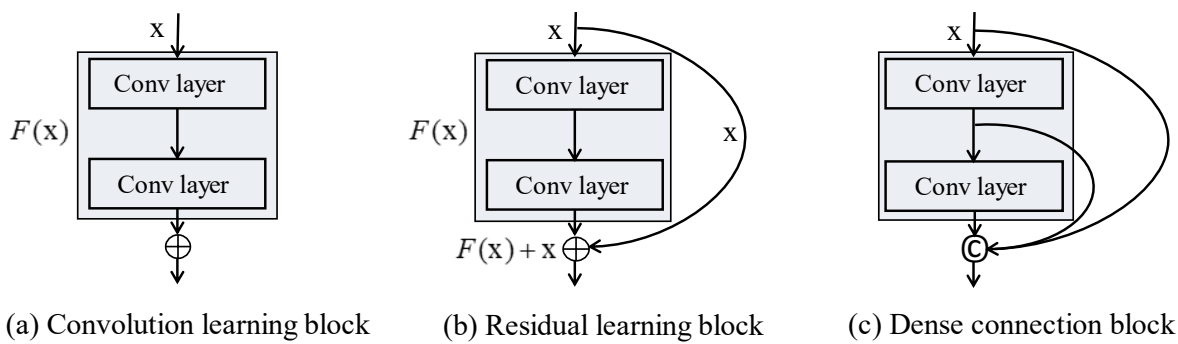

(a) Convolution learning block    (b) Residual learning block    (c) Dense connection block

*Figure 9 The traditional convolutional learning block, the residual learning block, and the dense connection block. Compared to the traditional convolutional learning block, the residual learning block is designed to learn residual features while preserving previous features through a skip connection. In contrast, the dense connection block is specifically designed to enhance feature propagation, promote feature reuse, and address the vanishing gradient problem by utilizing multiple skip connections. Unlike the residual learning block, which summarizes preceding feature maps, the dense connection block concatenates all feature maps from preceding layers. (For simplicity, activation and normalization layers are omitted.)*

Inspired by ResNet, a series of works have been proposed to modify the encoder of U-Net with residual learning blocks (see Figure 9 (b)) for enhancing segmentation performance. In [94], a weighted Res-UNet integrating the residual learning block into the encoder of U-Net, was proposed for retina vessel segmentation. These residual units facilitate gradient flow, enabling the training of deeper networks and improving model optimization. Building on this concept, SegResNet employs 3D ResNet-like blocks with group normalization for 3D MRI brain tumor segmentation, demonstrating its effectiveness in volumetric medical imaging [95]. Motivated by the success of multi-scale feature learning in Inception [96] and residual learning in ResNet, MultiResUNet was introduced for biomedical image segmentation, combining the strengths of both approaches [97]. A comprehensive review of residual learning in [93] further analyzed how to design effective and efficient residual blocks for various tasks.

Unlike ResNet, which employs skip connections for residual learning, DenseNet was introduced in [98] to maximize information flow and encourage feature reuse by utilizing multiple skip connections within a block (see Figure 9 (c)) by concatenating feature maps of all preceding layers. Building on its effectiveness in feature propagation and its efficiency in reducing the number of parameters, dense connection blocks were incorporated into the encoder of U-Net, resulting in the development of a novel hybrid densely connected U-Net (H-DenseUNet) [99]. This architecture combines a 2D DenseUNet for extracting intra-slice features with a 3D DenseUNet for aggregating volumetric contexts, achieving significant advancements for liver and tumor segmentation on CT volumes. Dense connections are also utilized in [100-102] for medical image segmentation.

Furthermore, the residual and dense connection blocks have been combined into the encoder of DR-VNet for retinal vessel segmentation, forming a Residual Dense-Net block (see Figure 10) [103]. The Residual Dense-Net block integrates a dense connection block and a residual learning block. The dense connection block effectively concatenates all preceding feature maps, ensuring no loss of information, while the residual learning block functions as a filter to extract and refine representative features.

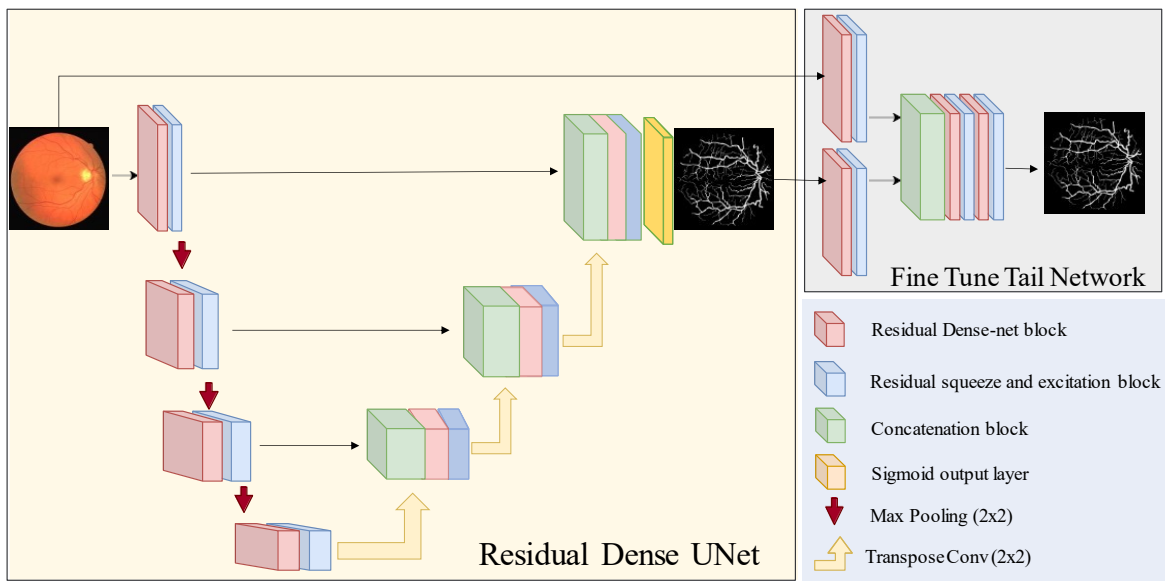

*Figure 10 Architecture of DR-VNet, which consists of a series of residual dense-net blocks in the encoder.*

Incorporating residual learning and dense connection blocks into the U-Net encoder enhances feature representation and improves final segmentation results. However, these architectures (such as ResNet and DenseNet) were originally designed for natural image classification tasks and may not be fully optimized for image segmentation. In particular, the substantial down-sampling operations in these designs can result in the loss of critical spatial information, which is essential for precise segmentation. To address this limitation, a Haar wavelet-based downsampling method was proposed in [91] , demonstrating its effectiveness in preserving crucial details when integrated into residual blocks for segmentation tasks. Additionally, various approaches, such as V-Net [16] and HRNet [25], continue to explore innovative skip connection strategies to further enhance the encoder's capability in biomedical image segmentation.

(2) *Variants of convolution operations in the encoder*

Since the success of VGG [92], the use of small (3×3) convolution filters with increasing layer depth has become the standard choice for image classification. This design reduces the number of training parameters compared to larger kernels while expanding the receptive field in a hierarchical manner. However, this approach is not always optimal for image segmentation, particularly in terms of balancing accuracy and computational efficiency. To overcome these limitations, two primary strategies have emerged for convolution operations in image segmentation (see Figure 11). The first strategy focuses on enlarging the receptive field within each layer. This is achieved through methods such as integrating Inception modules [104], employing deformable convolutions [105], or utilizing large convolutional kernels [106, 107]. The second strategy aims to enhance the efficiency of feature map learning by decomposing traditional convolution operations. Representative techniques include depth-wise-separable convolutions [108-110], group convolutions [14, 111], channel shuffling [112], and ghost modules [113]. These approaches collectively advance the trade-off between accuracy and efficiency in image segmentation tasks.

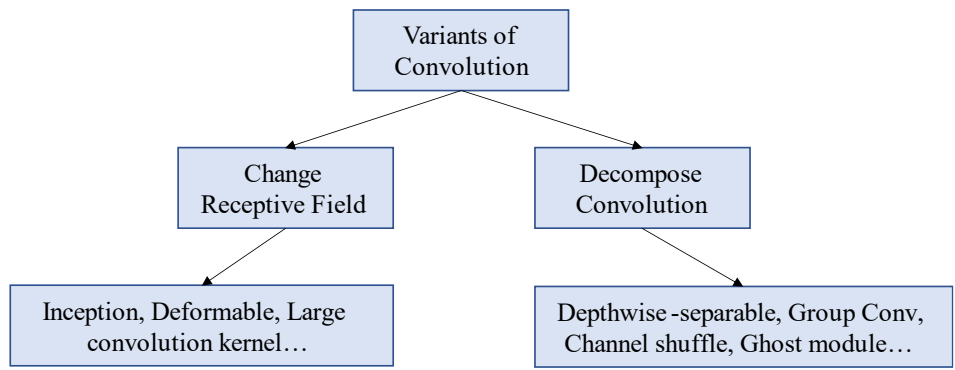

*Figure 11 Schematic of approaches for variants of convolution operations.*

In the Inception module, feature maps from preceding layers are passed into parallel convolutional layers with different kernel sizes and a max-pooling layer (see Figure 12). This design not only increases the width of each layer (the number of units at each level) but also enhances the network's depth, due to its ability to effectively process multi-scale feature maps. Building on these advantages, several U-Net-like architectures have integrated the Inception module into their encoders for medical image segmentation. Examples include MultiResUNet [97], multi-inception-UNET [114], and DENSE-INception U-Net [115], among others.

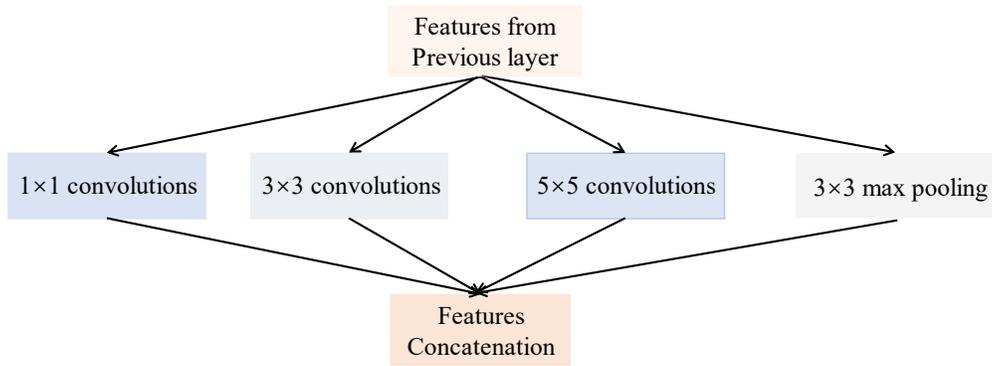

*Figure 12 The initial version of the Inception module.*

Deformable convolution was initially proposed in [105] to enhance the spatial sampling capability (receptive field) of standard convolution, offering greater flexibility in capturing spatially varying features (see Figure 13). Building on this concept, DUNet was developed for retinal vessel segmentation [116], incorporating deformable convolution into a U-shaped architecture. This integration allows the network to adaptively adjust the convolutional kernel shape to accommodate complex vessel structures. Similarly, in [117], deformable convolution was combined with spatial attention blocks within a U-Net framework to address variations in the shapes and sizes of different organs, further improving segmentation performance.

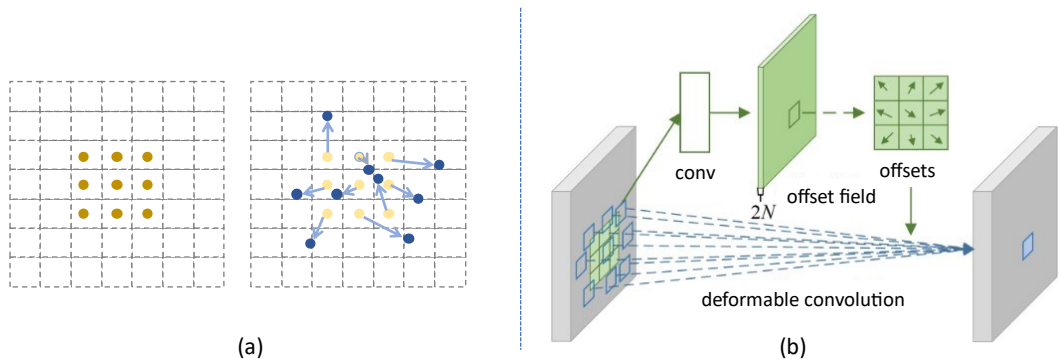

*Figure 13 Illustration of (a) a standard 3×3 convolution kernel (left) versus a 3×3 deformable convolution kernel (right). The arrows indicate learnable location offsets, denoting from where the value should come for the convolution operation. (b) The operation of 3x3 deformable convolution on a feature map. Unlike regular convolution, deformable convolution learns additional offsets to improve the spatial sampling capability of the standard convolution, enabling greater flexibility in capturing spatially varying features.*

In contrast to incorporating multi-scale kernels in the Inception module or varying kernel shapes in deformable convolution, one straightforward approach to enlarging the receptive field is to use larger convolutional kernels. However, this method inevitably increases the number of model parameters and may complicate training [92]. To address this, a Global Convolution Network (GCN) was proposed in [106] for image segmentation. GCN employs large one-dimensional kernels to expand the effective receptive field, enabling more efficient pixel-level classification. However, GCN overlooks the importance of multi-scale feature extraction. To overcome this limitation, SegNeXt [118] integrates global convolution into a multi-scale convolutional attention module, effectively combining multi-scale feature extraction with large kernels to enhance segmentation performance. In Figure 14, it illustrates the structure of global convolution and multi-scale convolutional attention block.

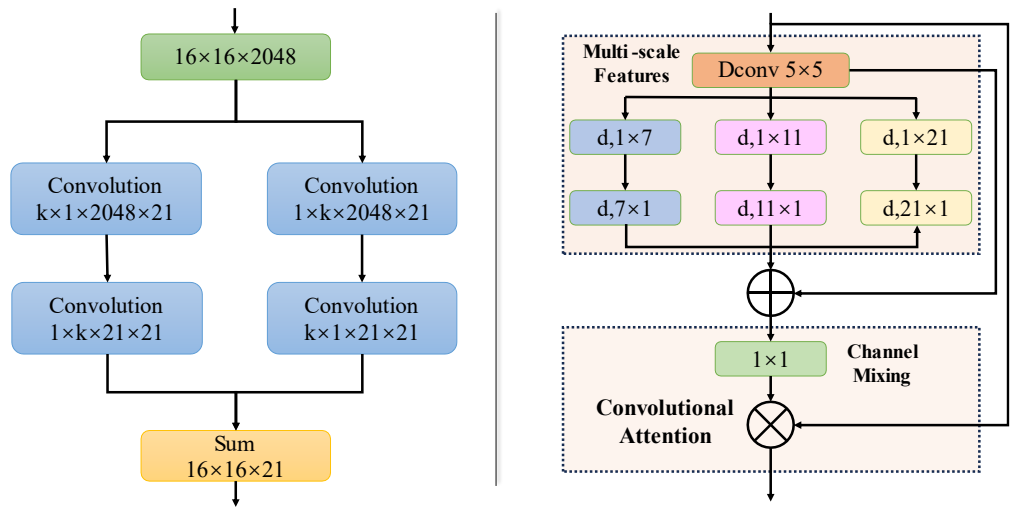

*Figure 14 The block of global convolution in GCN (left), and a variant of global convolution in SegNeXt (right).*

Compared to enlarging the receptive field in each convolutional layer of the encoder, decomposing convolution operations focuses on reducing computational overhead for deep neural networks. Depthwise separable convolution, introduced in MobileNet [108], decomposes a standard convolution into a channel-specific convolution (applying a single filter per input channel) followed by a 1×1 pointwise convolution, significantly reducing floating-point operations, parameters, and memory requirements. ResNeXt [111] extended this approach with grouped convolution, which divides feature maps into groups to increase cardinality and improve accuracy. When the number of groups equals the number of channels, grouped convolution simplifies to depthwise separable convolution. Building on these innovations, ShuffleNet [112] introduced channel shuffling within pointwise group convolution to further enhance efficiency. To address the memory and computational overhead still associated with the 1×1 pointwise convolution, GhostNet [113] introduced the Ghost module, which removes the pointwise convolution and employs skip connections to reuse feature maps. Figure 15 compares regular convolution, depthwise separable convolution, and the Ghost module.

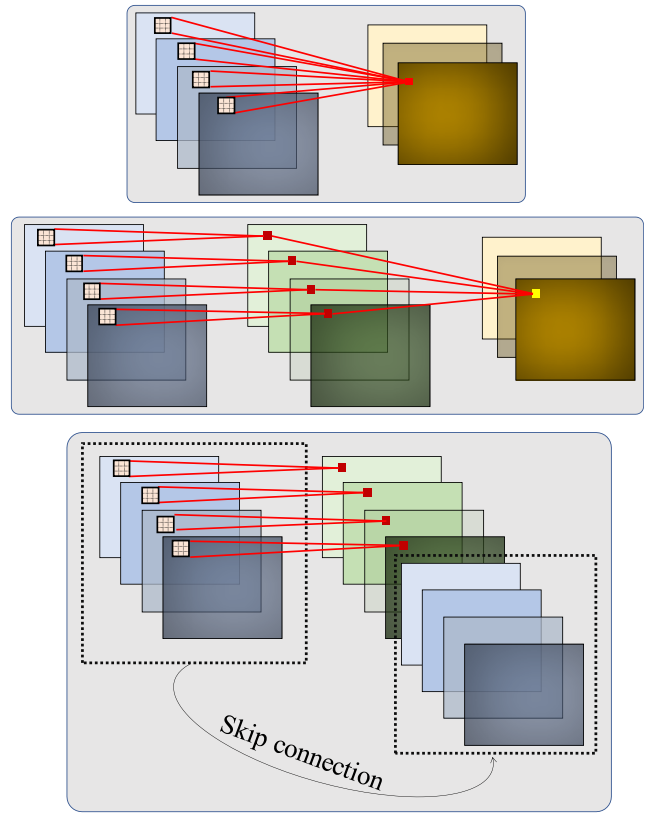

*Figure 15 Illustrations of Regular Convolution (Top), Depthwise Separable Convolution (Middle), and Ghost Convolution (Bottom). Regular convolution densely connects all input pixels. Depthwise separable convolution decouples spatial and channel-wise filtering. Ghost convolution leverages depthwise convolution and skip connections to reuse features efficiently.*

Such convolution-decomposing modules have also been applied to medical image segmentation, where the regular convolution in the encoder is typically replaced by one of these modules to reduce training parameters and computational costs while maintaining segmentation performance. For instance, MobileNetV2 [119] was employed as the encoder in a U-Net architecture for semantic segmentation of gastrointestinal cancer, leveraging depthwise separable convolutions for parameter-efficient learning in [120]. Similarly, ConvUNeXt [121] was developed for medical image segmentation, combining depthwise separable convolutions with large kernels to achieve a balance between parameter reduction and superior segmentation performance. Furthermore, group convolution, channel shuffling, and the Ghost module have been utilized in ResGANet [122], Group-PPM-Net [123], and GA-UNet [124], respectively, to enhance efficiency and accuracy in medical image segmentation tasks.

(3) *Attention mechanisms for convolution block in encoder*

In contrast to the above-mentioned methods, which aim to learn new representative features through skip connections or variable convolutional operations, the attention mechanism focuses on identifying the importance of preceding feature maps while restraining irrelevant information. This approach is based on the hypothesis that not all pixels in the feature maps contribute equally to the final object segmentation [125]. Generally, attention mechanisms used in convolutional neural networks can be categorized into three main types: channel or temporal attention, spatial attention, and self-attention. (Note that we exclude the novel self-attention-based Transformer architecture from this section, as Transformer-based methods are increasingly recognized as a distinct category in medical image segmentation research. These methods will be discussed in Section 3.5.)

In SENet [22], a novel Squeeze-and-Excitation block (see Figure 16) was introduced for image classification, designed to learn the importance of each feature map channel. The SE block comprises two key operations: squeeze and excitation. The squeeze operation reduces each channel to a single value using global average pooling, capturing global spatial information. The excitation operation generates an attention vector through fully connected layers and non-linear transformations, which is then used to recalibrate the feature map channels adaptively.

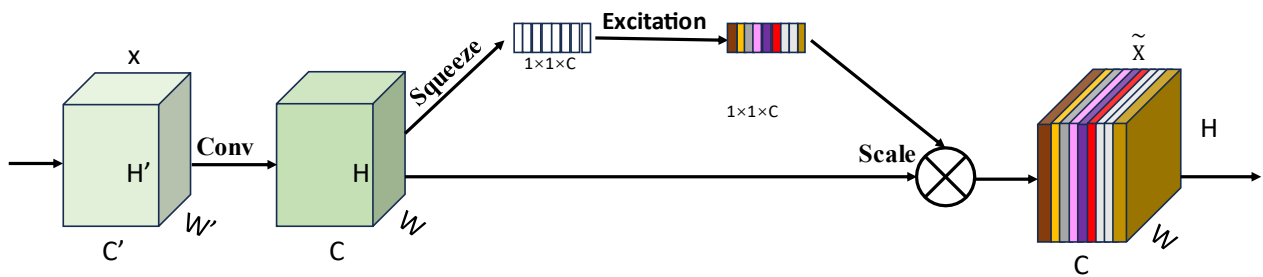

*Figure 16 A Squeeze-and-Excitation block.*

Extending this idea, the Convolutional Block Attention Module (CBAM) [23] was proposed to refine features adaptively by learning attention maps sequentially along both channel and spatial dimensions. Similarly, DANet [126] integrates position attention (spatial dimension) and channel attention (channel dimension) modules in parallel, enabling the fusion of local features with global dependencies for improved image segmentation (see Figure 17). Unlike previous methods that learn a spatial attention map and apply it uniformly to all preceding feature maps, self-attention mechanisms focus on capturing long-range dependencies for each pixel within individual feature maps. NonLocal [127] and CCNet [128] leverage self-attention to model contextual information effectively, addressing the inherent locality limitations of traditional convolutional operations.

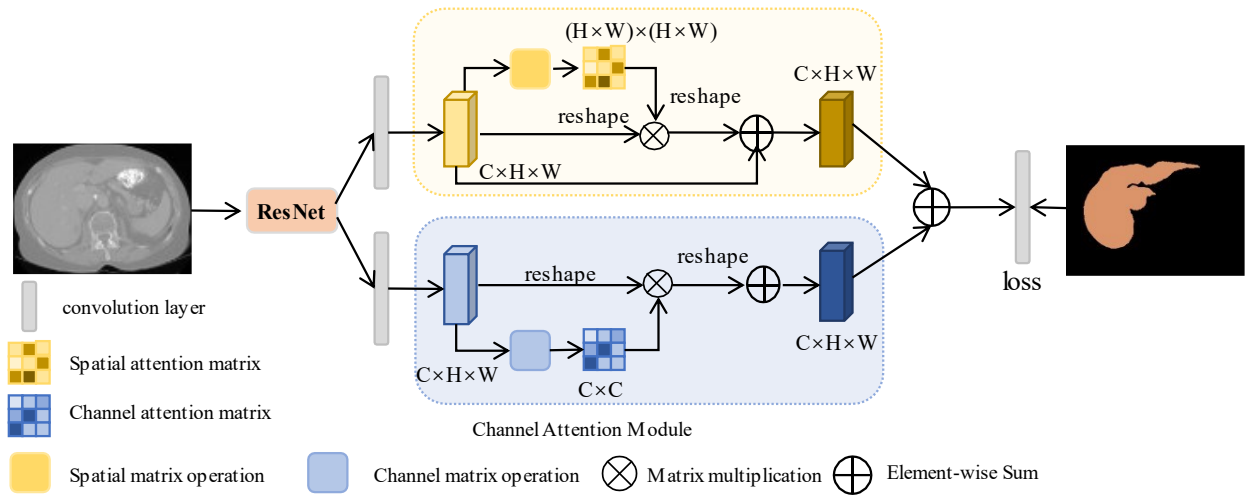

*Figure 17 The Architecture of Dual Attention Network.*

In medical image segmentation, attention mechanisms such as *channel-wise* and *position-wise spatial attention* are frequently integrated into encoder architectures [85]. For instance, LAEDNet [129] employs lightweight residual Squeeze-and-Excitation blocks in the encoder to segment ultrasound images, effectively capturing channel-wise relationships between feature maps. Similarly, spatial and channel attention mechanisms in [130] serve as auxiliary modulation blocks within the U-Net encoder, adaptively emphasizing discriminative features while suppressing irrelevant information. Building on this, CPCANet [131] introduces an efficient channel-prior convolutional attention mechanism that dynamically allocates attention weights across both channel and spatial dimensions, leveraging a multi-scale depthwise convolutional module.

Furthermore, self-attention-based blocks have been incorporated into U-Net encoders to effectively aggregate global contextual information, enhancing feature representation and segmentation accuracy. For example, Non-Local U-Net [132] integrates global aggregation blocks based on self-attention, significantly enhancing biomedical image segmentation. A Triple Attention Network (TA-Net) [133] leverages the attention mechanism to simultaneously capture global contextual information across the channel, spatial, and feature internal domains. Specifically, a channel with a self-attention encoder block is integrated into the U-Net encoder to learn the long-range dependencies of pixels, enhancing feature representation and segmentation performance. Overall, these attention mechanisms—channel, spatial, and self-attention—are typically employed as plug-ins within convolutional blocks, effectively emphasizing significant feature maps or spatial regions while suppressing less relevant information.

## 3.2 Improving receptive field in the bottleneck

In U-Net, the bottleneck serves as a crucial bridge between the encoder and decoder, compressing and extracting essential information from the input data [36]. As the features have been progressively downsampled through the encoder, the bottleneck benefits from a large receptive field, enabling it to capture rich contextual information and extract high-level semantic representations. This condensed representation facilitates the decoder in accurately localizing and identifying the target structures for segmentation. Typically, the bottleneck contains numerous feature maps with low spatial resolution (e.g., 1024 feature maps in the vanilla U-Net). However, the fixed convolution kernels in U-Net constrain its receptive field, limiting its ability to capture objects of varying sizes or incorporate large contextual regions. This limitation can result in suboptimal representative features in the bottleneck for accurately segmenting objects across diverse scales. To address this, various methods have been proposed to enhance the bottleneck's functionality, primarily by enlarging the receptive field to extract multi-scale features and employing attention mechanisms for adaptive recalibration of important features.

(1) Enlarge the receptive field in the bottleneck

Three primary approaches are commonly employed to enlarge the receptive field in a convolutional neural network, particularly within the bottleneck stage for image segmentation. The first involves utilizing various convolution kernels to extract multi-scale feature maps, which are subsequently concatenated or summarized to capture contextual information across multiple scales (see Figure 18 (a)). Representative examples include the Inception family [96, 104, 134]. The second approach employs subsampling operations to produce pyramidal feature maps, as demonstrated in Spatial Pyramid Pooling (SPP), initially introduced in SPP-net for object detection [135] and later extended to image segmentation in the Pyramid Scene Parsing Network (PSPNet) [136] and the Pyramid Attention Network (FPN) [137] (seeFigure 18 (b)). The third approach, atrous spatial pyramid pooling (ASPP), uses dilated convolutions with varying dilation rates to effectively capture multi-scale contextual information [20, 138] (see Figure 18 (c)).

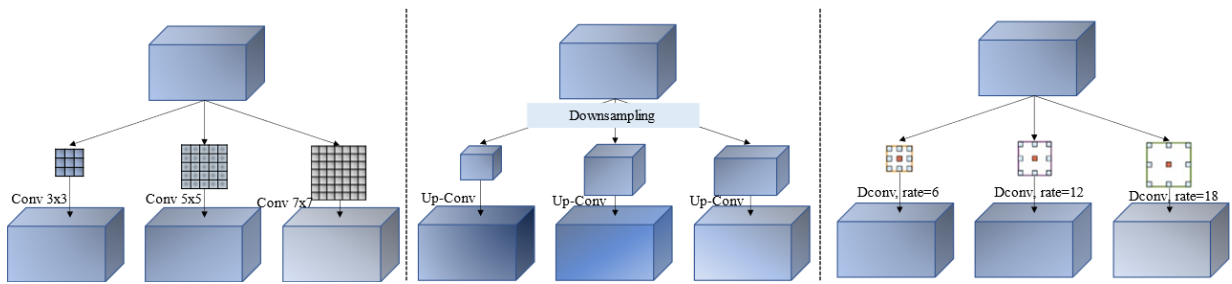

*Figure 18 Three approaches for multi-scale feature extraction to enlarge the receptive field. (a) Utilizing convolution kernels of various sizes, like the Inception module; (b) Employing down-sampling to generate multi-scale features; (c) Using dilated convolutions by inserting zeros between kernel elements.*

One limitation of Inception-based methods, which enlarge convolutional kernel sizes for multi-scale feature extraction (Figure 18 (a)), is the significant increase in model parameters, leading to challenges in training deep neural networks. To address this issue, pointwise convolution and residual connections were introduced in Inception-v4 [134], and depthwise separable convolution was employed in Xception [110]. A variant of the Inception module for multi-scale feature extraction is Res2Net [139] (see Figure 19). Res2Net constructs hierarchical residual-like connections within a single residual block to represent multi-scale features at a granular level, effectively increasing the receptive field range for each network layer. Unlike Inception-like methods, it employs a consistent convolution kernel size combined with hierarchical skip connections to extract multi-scale features in the proposed Res2Net module. Several studies have adapted Inception-like modules for medical image segmentation by integrating them into the encoder and bottleneck, demonstrating improved performance in applications such as brain tumor, liver tumor, prostate, and vessel segmentation [114, 115, 140-142].

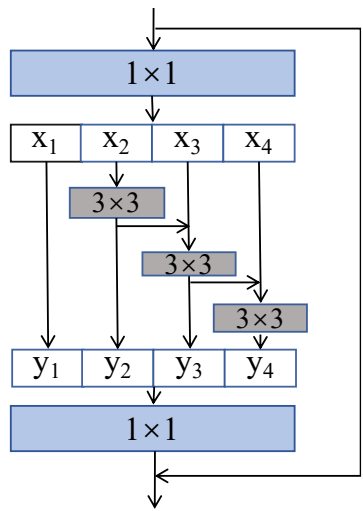

*Figure 19 Schematic of the Res2Net module. It extracts multi-scale features within a residual block using a hierarchical convolutional approach.*

In contrast to using varying convolution kernel sizes, pyramidal architectures employ down-sampling (pooling) operations to generate feature maps at multiple scales. Convolutions with a uniform kernel size are then applied to these multi-scale feature maps, effectively extracting contextual information (Figure 18 (b)). PSPNet [136] introduces a pyramid parsing module (PPM) designed to capture sub-region representations at different scales. This module utilizes pooling operations with varying strides to produce multi-scale feature maps, followed by parallel convolution operations applied to these maps. Finally, up-sampling and concatenation layers combine the features into a unified representation that incorporates both local and global contextual information (Figure 20). Subsequently, the feature representation is fed into a convolution layer to obtain the final per-pixel prediction. In LeViT-UNet [30], multi-scale features from different layers of a Transform module are upsampled to the same spatial resolution and concatenated for contextual information fusion within the bottleneck. In [143], a 3D pyramid pooling module was integrated as a bottleneck into each skip connection of U-Net for prostate segmentation in MRI. This module performs multifaceted feature extraction on feature maps from various convolutional layers, helping to concentrate multi-scale information of the segmented structures and thereby enhancing segmentation accuracy.

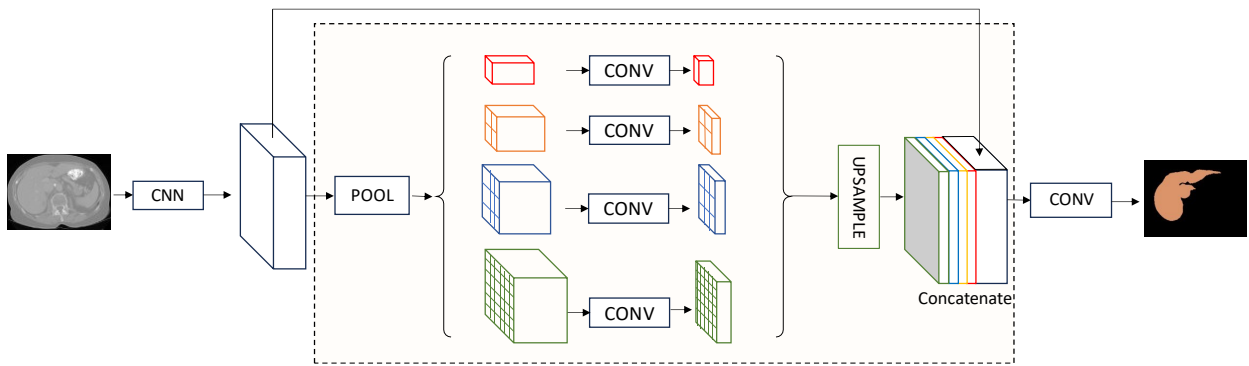

*Figure 20 Pyramid Pooling Module (dashed box) in PSPNet. The module performs pooling at multiple scales to generate feature maps, which are then refined via convolutions. These multi-scale features are upsampled and concatenated to effectively fuse contextual information. The fused representation is subsequently passed through a convolutional module to produce the final prediction mask (orange).*

Another effective method to enlarge the receptive field without adding extra parameters is atrous convolution (also known as dilated convolution) [20, 144]. Unlike standard convolution, it introduces "holes" (zeros) between filter elements, thereby expanding the kernel's coverage over a broader area. This technique allows feature responses to be computed at any desired resolution on feature maps, eliminating the need for subsampling or large kernels that would otherwise add computational and parameter overhead (see Figure 21).

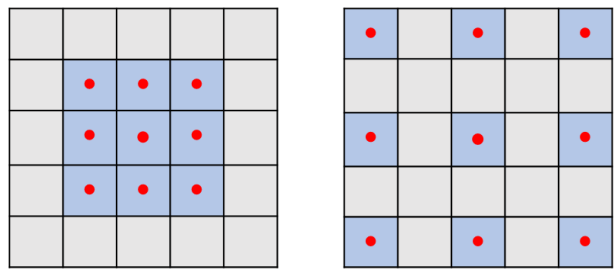

*Figure 21 Standard convolution (left) versus atrous convolution with a stride rate of 2. Atrous convolution introduces holes between kernel elements, enabling an enlarged receptive field without increasing training parameters.*

Atrous spatial pyramid pooling (ASPP) extends this concept by probing input feature maps at multiple dilation rates in parallel, effectively capturing contextual information across various scales and significantly enhancing segmentation performance. For instance, Deeplabv3+ ASPP [138] combined dilation with depthwise separable convolution for integration into an encoder-decoder structure, resulting in a faster and more robust network. Specifically, ASPP is embedded into the bottleneck of the encoder-decoder architecture, maximizing its ability to extract and utilize multi-scale contextual information (see Figure 22).

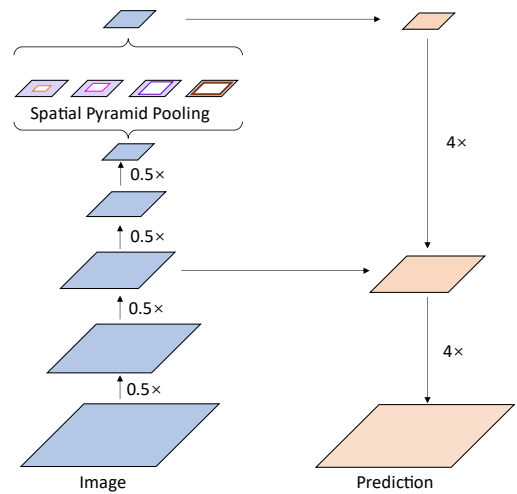

*Figure 22 Encoder-decoder structure with ASPP.*

Inspired by the concept of ASPP for multiscale feature extraction in object segmentation, its application has been extensively explored in the bottleneck of U-Net architectures to enhance multiscale representation. In ResUNet++ [145], ASPP acts as the bridge between the encoder and decoder, effectively expanding the field of view for filters, thereby improving polyp detection and segmentation in colonoscopic images. Similarly, in COPLE-Net [146], ASPP is embedded into the bottleneck of the encoder-decoder structure, comprising four parallel layers of dilated convolutions with dilation rates of 1, 2, 4, and 6. The outputs of these layers are concatenated to capture features across multiple scales, enabling effective segmentation of both small and large COVID-19 pneumonia lesions in CT images. This highlights the versatility and efficacy of ASPP in diverse medical imaging applications.

In another approach, a Joint Classification and Segmentation (JCS) system [147] was proposed for COVID-19 lesion segmentation. This system integrates a classification branch for generating explainable results and a segmentation branch for precise lesion delineation. Two Grouped Atrous Modules (GAM) are incorporated into the bottleneck to extract representative feature maps with larger receptive fields. The process begins with 1×1 convolution to expand feature map channels, followed by dividing the feature maps into four groups. Atrous convolutions with varying dilation rates are then applied to these groups, resulting in enriched feature maps with diverse receptive fields.

ASPP has also been integrated into U-Net variants for various medical imaging tasks, including breast lesion segmentation on mammograms [148], thyroid nodule segmentation on ultrasound images [149], and liver segmentation on CT scans [150]. Notably, DoubleU-Net [151] leverages two stacked U-Net architectures for medical image segmentation across modalities such as colonoscopy, dermoscopy, and microscopy. This architecture integrates two ASPP modules at the bottlenecks, effectively capturing contextual information and achieving superior segmentation accuracy. Results demonstrate that DoubleU-Net significantly improves upon the standard U-Net model, producing more precise segmentation masks (see Figure 23).

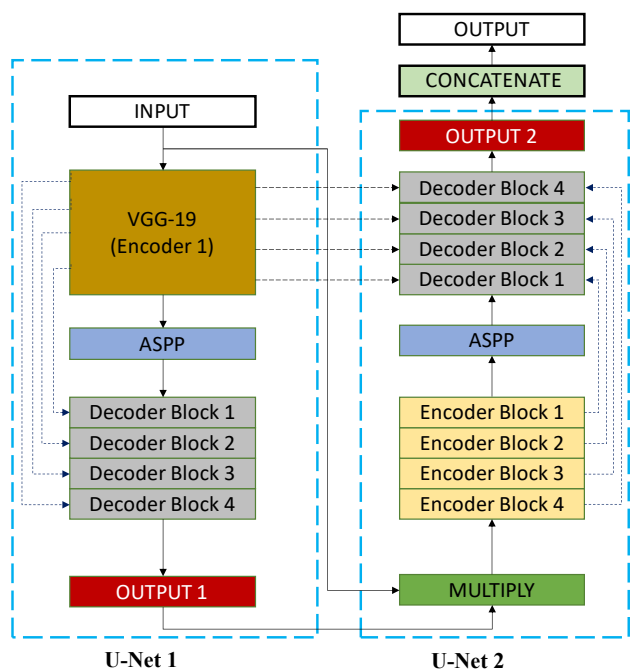

*Figure 23 The Block diagram of the proposed DoubleU-Net architecture, where two ASPPs are utilized as bottlenecks for resampling features at multi-scale levels.*

(2) Attention mechanisms in the bottleneck

In contrast to pyramidal architectures in the bottleneck, several studies have incorporated attention mechanisms into the bottleneck of U-Net to enhance feature representation. For instance, DEAU-Net [152] employs a combination of a residual attention block and a channel attention block in the bottleneck. Within the residual attention block, input feature maps are divided into two groups. The first group utilizes 1×3 and 3×1 convolution to learn spatial attention for residual features, while the second group integrates feature maps into residual features. Subsequently, a channel attention mechanism based on selective kernel convolution [153] is applied to achieve comprehensive extraction and fusion of full-scale information at the channel level.

Similarly, SECA-Net [154] introduces two modules in a sequential manner within the bottleneck: The Squeeze-and-Excitation Multi-Scale Atrous Convolution (SEMAC) module and the Squeeze-and-Excitation Spatial-Channel Attention (SESCA) module. These modules refine deep features from the encoder path, enabling the capture of spatial-channel-aware multiscale features for enhanced segmentation.

MA-Net [155] introduces a position-wise attention block in the bottleneck to model feature interdependencies in spatial dimensions, capturing global spatial dependencies between pixels (see Figure 24). The process begins by splitting the feature maps into three groups using 1×1 convolution. A spatial attention map is then generated through reshaping, multiplication, and softmax operations. Concurrently, the third group of feature maps is reshaped and multiplied with the attention map to produce weighted feature maps. Finally, the original feature maps are combined with the weighted feature maps for refined representation. In addition, spatial attention mechanisms have been successfully integrated into other U-Net variants, such as SA-UNet [156], CANet[157], CS$^2$-Net [158] and CHWS-UNet [159], further demonstrating the effectiveness of attention mechanisms in enhancing feature representation within the bottleneck.

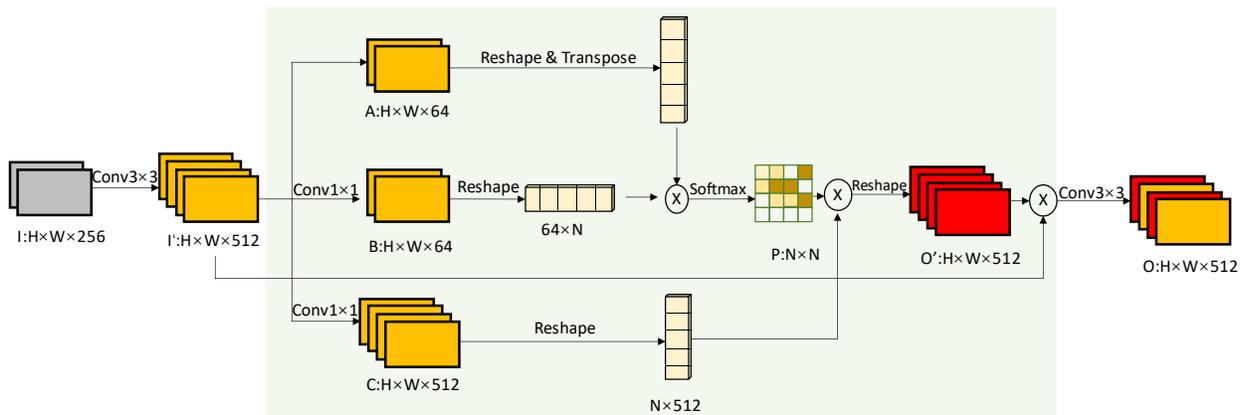

*Figure 24 Structure of position-wise attention block (with green background)*

## 3.3 Multi-Scale features in the decoder

The decoder in segmentation networks is designed to progressively up-sample feature maps to the desired resolution, ultimately generating the final segmentation masks for each object. Many decoders adopt architectures similar to their corresponding encoders, as seen in SegNet [51], LinkNet [62], DeConvNet [160], U-Net, V-Net [16], and RA-UNet [161]. While U-Net-like methods incorporate feature maps from multiple layers, they often face challenges in effectively integrating multi-scale information within the decoder. This limitation can significantly impact their ability to accurately segment objects with varying shapes and sizes.

Generally, two primary approaches have been proposed to address multiscale feature integration in the decoder. The first approach focuses on multiscale feature fusion within the decoder, where feature maps from different scales are combined to produce the final segmentation results (see Figure 25(a)). The second approach involves utilizing multiscale features to predict segmentation masks at multiple scales, which are then ensembled to generate the final segmentation mask (see Figure 25(b)).

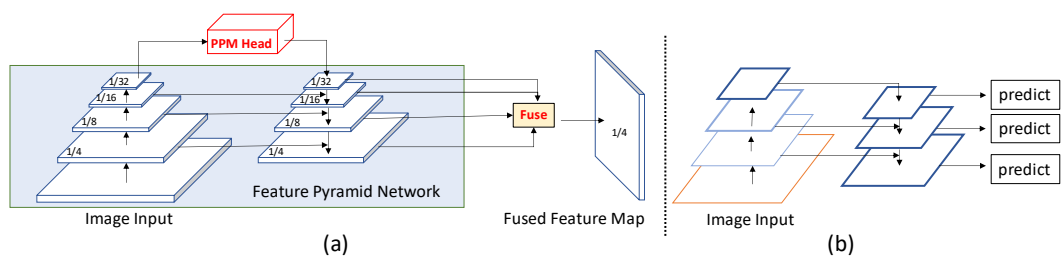

*Figure 25 Two strategies for handling multiscale features in the decoder. (a) Multiscale feature maps are fused to produce the final segmentation result. (b) Multiscale feature maps are used to predict segmentation masks independently, which are then ensembled for the final output.*

(1) Multi-scale feature or context extraction and fusion in the decoder

The fusion of multi-scale features in the decoder plays a crucial role in achieving accurate segmentation results. Multi-scale features capture multi-level contextual information, effectively describing the pixel-value characteristics and inter-pixel relationships of objects. In UPerNet [162], the Feature Pyramid Network (FPN) [71] is utilized as the encoder, paired with a pyramid pooling module in the bottleneck. Within the decoder, multiscale feature maps are systematically fused, starting from low-level high-resolution feature maps and extending to high-level low-resolution feature maps from all decoder layers. To unify feature maps of varying scales, bilinear interpolation is employed to adjust all features to the spatial resolution of the low-level feature maps. A convolutional layer is then applied to integrate features from different levels and reduce channel dimensions, ensuring efficient and effective feature fusion (see Figure 25(a)).

To establish long-range relationships, multi-scale features are often combined with self-attention mechanisms. However, the computational overhead of self-attention in non-local networks presents a significant challenge. To address this, an Asymmetric Fusion Non-local Block (AFNB) (see Figure 26) was introduced in ANN [163] and implemented within the decoder for semantic segmentation. The AFNB integrates a pyramid pooling module into the non-local block, effectively reducing computational costs while maintaining segmentation performance. By leveraging pyramid pooling, the AFNB fuses multi-scale features while simultaneously capturing long-range dependencies.

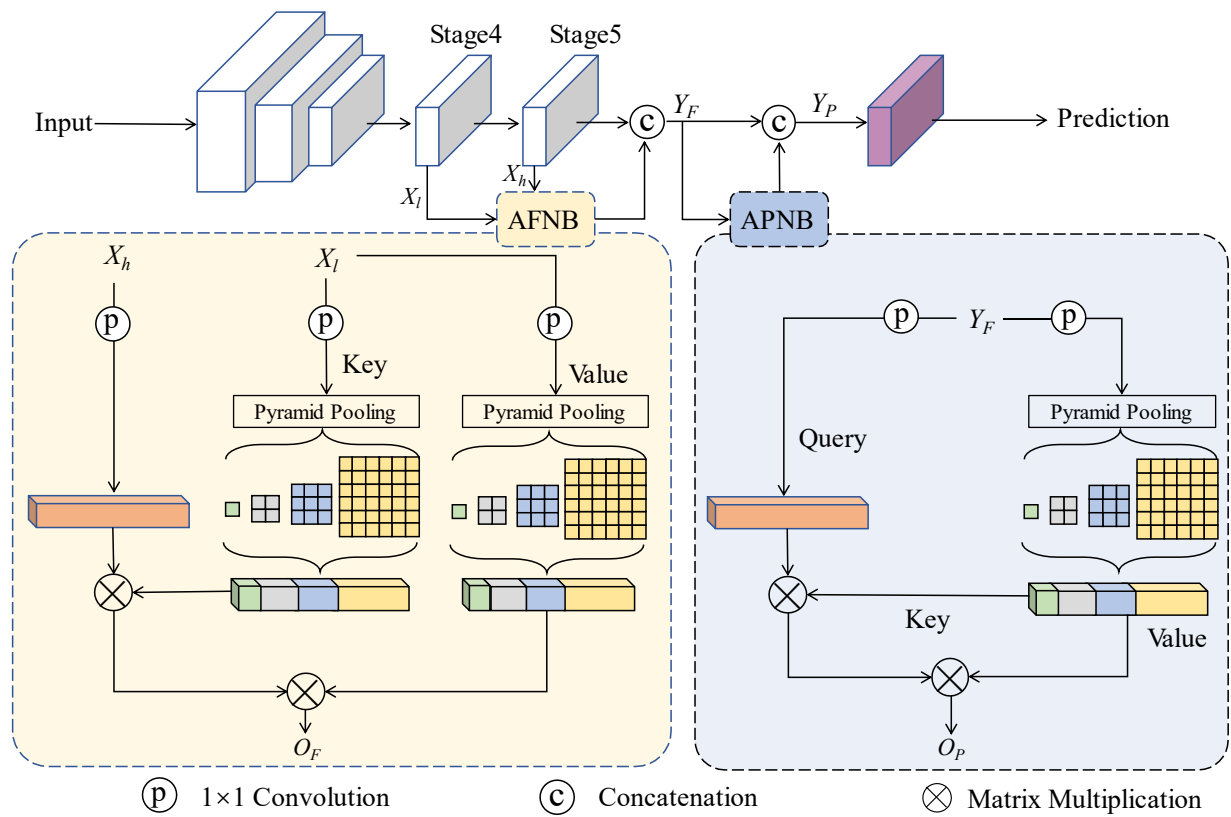

*Figure 26 Overview of the asymmetric non-local neural network. In the implementation, the key and value branches in the AFNB share the same 1×1 convolution and sampling module, significantly reducing the number of parameters and computational complexity without sacrificing performance.*

Unlike the multi-scale feature extraction and fusion in spatial level, several works focus on relational context level by aggregating multi-class or multi-semantic representation for segmentation, like ACFNet [164], OCR [165], DANet [126] and its related work [166-168]. Particularly, an object context aggregation scheme is proposed in OCNet [169], aiming to enhance the importance of object information. It uses an interlaced spares self-attention method and integrates a pyramid pooling module for multi-scale object context information extraction and fusion in the decoder (see Figure 27).

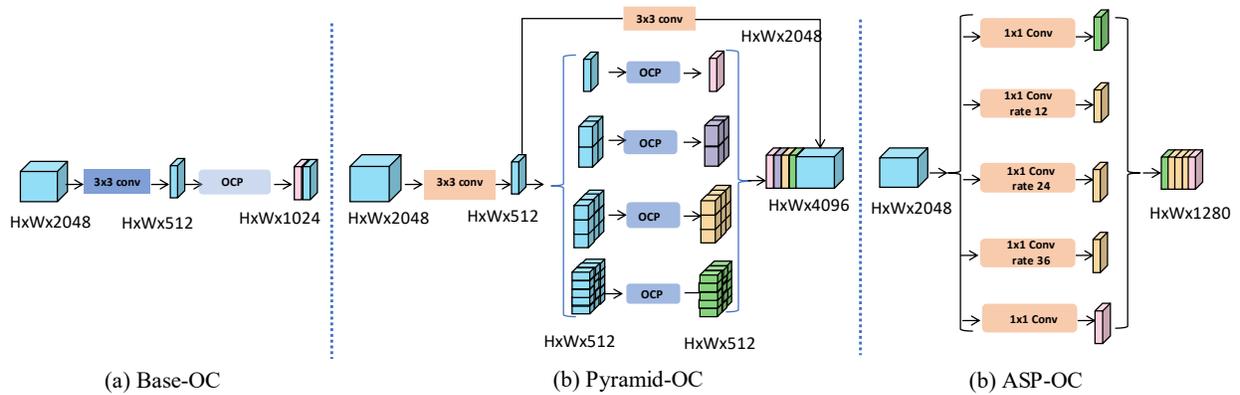

*Figure 27 Schematic of the modules of OCNet. The Object context pooling is integrated into a pyramidal structure with varying spatial scales.*

Likewise, multi-scale features and semantic information are also explored in medical image segmentation. In EMCAD [170], an efficient multi-scale convolutional attention decoder is introduced to optimize both performance and computational efficiency for medical image segmentation. To be specific, a multi-scale depthwise convolution block combines channel, spatial, and grouped (large-kernel) gated attention mechanisms to capture multi-scale information and intricate spatial relationships between feature maps. MERIT [171] introduces a cascaded attention-based method in the decoder for refinement of the multi-stage features. Particularly, two decoders across different window sizes with cascaded skip connections are used for enhancing the model capacity to capture multi-scale features critical for medical image segmentation. In FFUNet [172], a feature fusion module is present and integrated into the decoder, aiming to alleviate the ambiguous semantic information between the encoder and decoder. In a nutshell, we found multi-scale features are usually combined with attention mechanisms for effective feature fusion in the decoder for medical image segmentation. One of the possible reasons is the difference in multi-level semantic information from multi-scale features.

### (2) Predict segmentation masks at multiple scales after the decoder

Another effective strategy for utilizing the hierarchical structure of the decoder is to *predict segmentation masks at **multiple scales*** and fuse them to produce the final output. Originally developed for object detection, Feature Pyramid Networks (FPN) [91] exploit semantic feature maps across all decoder scales (see Figure 25(b)), achieving state-of-the-art performance on the COCO detection benchmark [173] with minimal additional complexity. Subsequently, FPN has been successfully adapted for natural image segmentation [102–104] and extended to medical image segmentation tasks [105–107]. For instance, RFPNet [105] introduces a reorganization feature pyramid network that constructs semantic features across multiple scales and levels (see Figure 28). This approach employs multi-branch feature decoders to generate a series of feature pyramids, which are subsequently processed through a feature pyramid reorganization module that stacks feature maps at the same scale. The final segmentation result is achieved by fusing the multi-scale weighted prediction masks. Similarly, in [108], a feature pyramid structure combined with a level-aware attention module is proposed to extract and refine multi-level features, demonstrating effectiveness in meningioma segmentation.

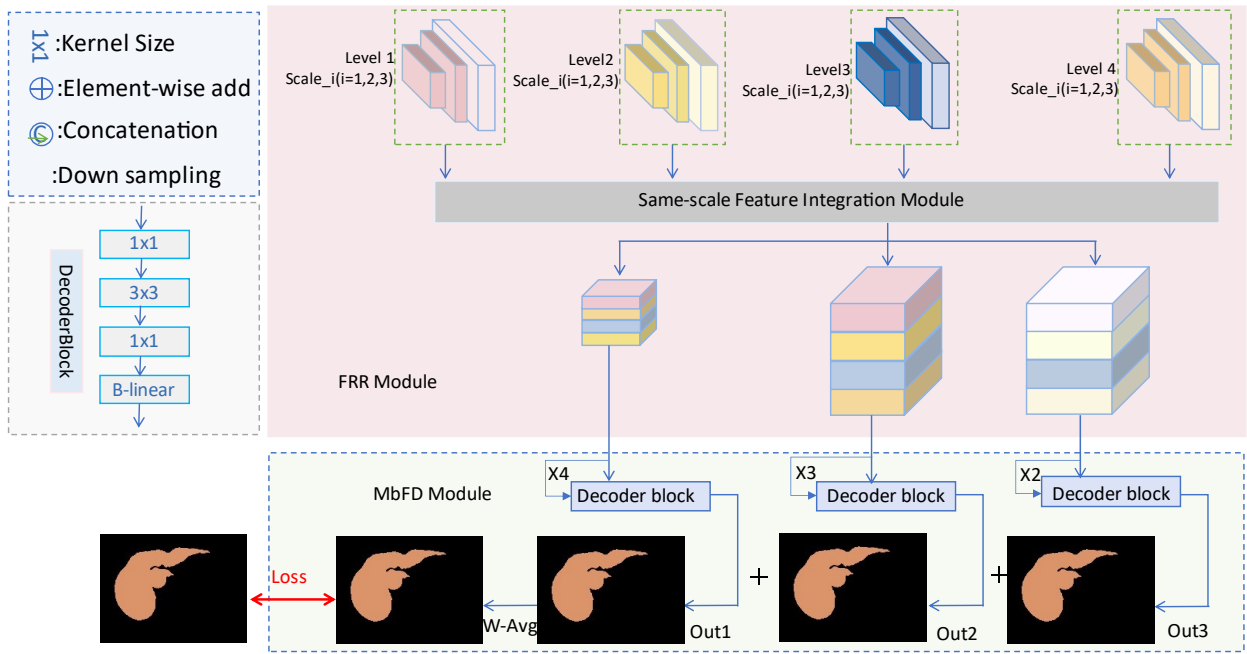

*Figure 28 Illustration of the reorganization feature pyramid network.*

Coupled with multi-scale prediction, the deeply supervised learning strategy [174] is widely adopted during the training process to address optimization challenges and accelerate convergence. This approach enhances the network's discriminative capability by introducing auxiliary supervision signals (loss functions) at intermediate layers, alongside the final output layer. By incorporating these additional supervisory signals, the model ensures that both the final output and intermediate feature representations contribute to the learning process, thereby improving training efficiency and effectiveness. For example, a 3D deeply supervised network was proposed in [175, 176] for volumetric medical image segmentation, effectively addressing challenges such as complex anatomical environments, optimization difficulties in 3D networks, and limited training samples. Additional advancements leveraging deeply supervised mechanisms are detailed in related works [177-180].

### 3.4 Effective spatial detail transition with the skip connection

The skip connections in FCNs and U-Nets, while enhancing feature integration, are limited by their inability to capture long-range dependencies, susceptibility to propagating noise, and potential misalignment during feature fusion, which can affect segmentation accuracy and robustness. Generally, there are two directions to address these problems to enhance the segmentation performance: (1) to integrate attention mechanisms into skip connections, and (2) multi-scale feature fusion with skip connections.

(1) Attention in the skip connection

In Attention U-Net [181, 182], an additive attention gate (illustrated in Figure 29) is proposed and integrated into the skip connections of U-Net to suppress irrelevant regions and emphasize areas of interest. Specifically, for three-dimensional feature maps with multiple channels, coarse-scale feature maps from the bottleneck or decoder outputs are transformed using 1×1×1 point-wise convolutions along the channel dimension. Concurrently, feature maps from the encoder are processed with point-wise convolutions and down-sampled. These two sets of feature maps are subsequently combined through addition. A series of operations, including ReLU activation, point-wise convolution, sigmoid activation, and linear interpolation, is then applied to generate a three-dimensional attention map. This attention map is finally multiplied with the original feature maps from the encoder to suppress task-irrelevant information. Moreover, convolution-based channel and spatial attention mechanisms have also been widely adopted and integrated into the skip connections of U-Net-based models for medical image segmentation. Examples include CA-Net [183], SCS-Net [184], MAD-UNet [185], which leverage attention mechanisms to enhance discriminative representations by effectively utilizing multi-granularity features from both the encoder and bottleneck/decoder.

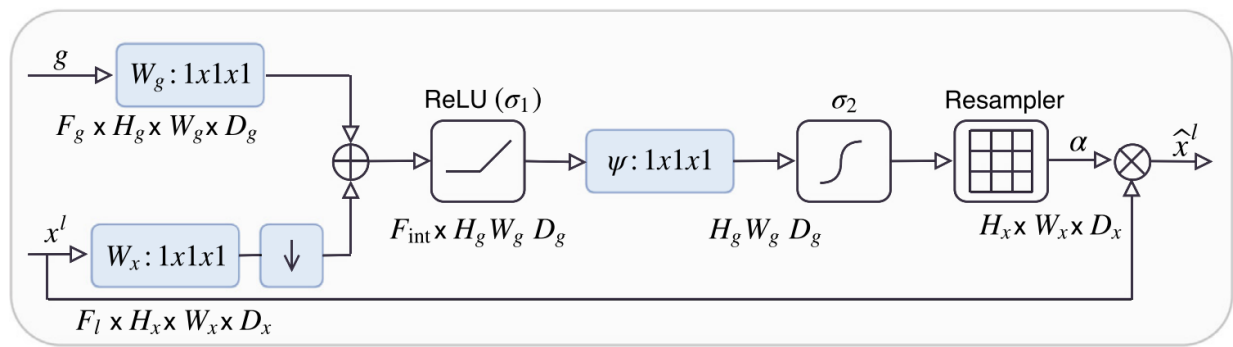

*Figure 29 Schematic of addition attention gate (AG). The feature maps from the encoder are recalibrated with an attention map computed with AG.*

Additionally, self-attention and cross-attention mechanisms have been incorporated into skip connections to fuse features from the encoder and decoder, capturing long-range contextual interactions and spatial dependencies[1]. In U-Transformer [186], self- and cross-attention blocks from Transformers are integrated into both the bottleneck and skip connections, with cross-attention in the skip connections enabling fine spatial information recovery by filtering out non-semantic features. Similarly, CoTr [187] employs deformable Transformer blocks [188] to establish long-range dependencies on all multi-scale features from the encoder. Variants of self-attention have also been adapted to enhance skip connections in medical image segmentation models. Examples include the channel-wise cross-fusion Transformer in UCTransNet [189], the enhanced Transformer context bridge in MISSFormer [190], and the Swin Transformer in SwinCross [191], all of which fuse multi-scale features to capture global contextual relationships and emphasize important spatial information. Figure 30 illustrates two approaches for integrating attention mechanisms into skip connections for medical image segmentation, highlighting the use of attention extraction on either single-scale or multi-scale features.

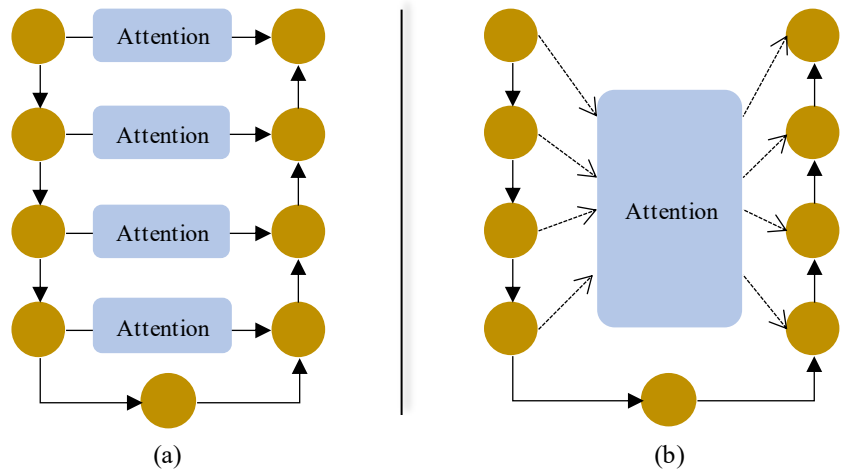

*Figure 30 Two approaches for integrating attention mechanisms in the skip connections. The yellow circle denotes a Transformer block, while the dashed arrow represents a skip connection that bypasses features from the encoder.*

(2) Multi-scale in the skip connections

Multi-scale features are widely utilized in skip connections to extract richer semantic information, thereby improving segmentation performance. This approach leverages the hierarchical structure of encoders, which inherently generate multi-scale feature maps. A notable example is UNet++ [192], designed to address two major limitations of U-Net and FCN: (1) the optimal network depth is not predetermined, necessitating exhaustive architecture searches or the ensembling of models with varying depths, both of which are computationally intensive; and (2) skip connections in the original U-Net transfer same-scale feature maps directly from the encoder to the decoder (seeFigure 31(a)), constraining multi-scale feature fusion. To address these challenges, the authors of UNet++ introduced a redesigned architecture that incorporates additional decoders and dense skip connections. This design unifies multiple shallow U-Net architectures into a single framework, enabling the united assembling of U-Nets with varying depths and promoting effective multi-scale feature fusion (see Figure 31(b)). In contrast to UNet++, which nests additional decoders but still employs single-scale feature transfers via skip connections, UNet3+ [193] achieves more comprehensive feature integration by utilizing full-scale skip connections. Specifically, feature maps from all encoder levels are concatenated and passed to each decoder layer, fully leveraging multi-scale features to enhance segmentation performance across organs of varying sizes.

[1] For completeness, this paragraph highlights representative Transformer-based methods that integrate attention into skip connections, while Section 3.5 examines complete segmentation architectures, including both Transformer-based and convolution–Transformer hybrids.

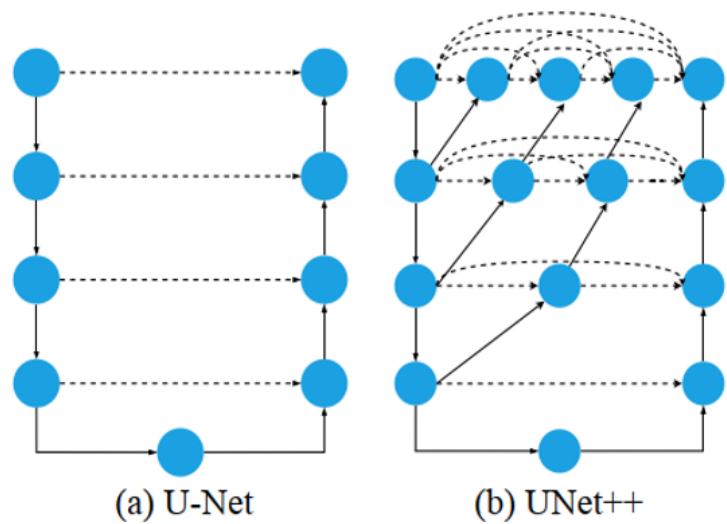

*Figure 31 The schematic of (a) U-Net and (b) UNet++. The blue dots indicate convolution-based modules used for feature extraction.*

Building on the foundation of UNet++, subsequent works have incorporated attention mechanisms and multi-scale fusion modules. Examples include Attention UNet++ [194], tailored for liver segmentation, as well as MSNet [195] and MS2Net [196], which focus on polyp segmentation. Additionally, pyramid-based and atrous spatial pyramid-based modules have been embedded into skip connections to further capture and merge multi-scale information for improved segmentation outcomes. Examples include CPFNet [197], SCS-Net [184], DiSegNet [198], HMEDN [199] among others.

## 3.5 Transformer-based and convolution-transformer hybrid methods

While FCNs and U-Net capture local and global context to some extent through skip connections, their ability to model long-range dependencies and complex contextual relationships is limited compared to more advanced architectures like Transformers [30]. As for segmentation, understanding the semantic relationships plays an essential role in the success of medical image segmentation. To this end, there are many works that explored Transformers for medical image segmentation. The main variations in terms of encoder-decoder architecture are illustrated in Figure 32.

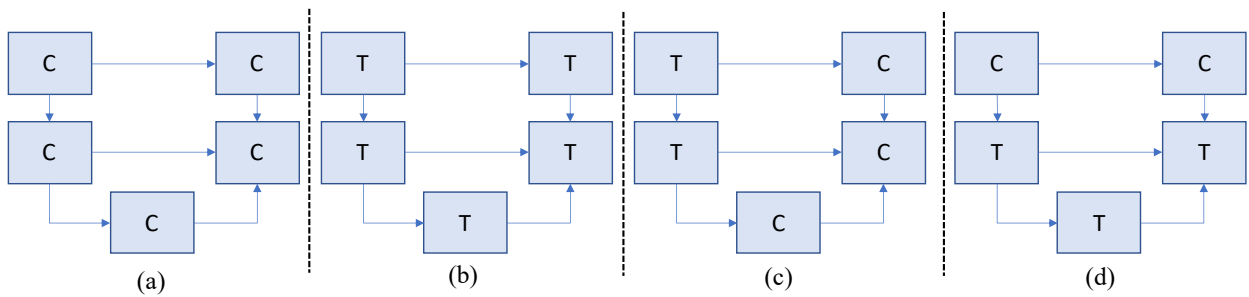

*Figure 32. Four architectural variants for modifying the encoder and decoder in segmentation networks. (a) A fully convolutional architecture, such as U-Net. (b) A pure transformer-based architecture, where both the encoder and decoder are composed entirely of transformer blocks, like SETR. (c) A hybrid design with transformer blocks in the encoder and convolutional blocks in the bottleneck and decoder, as seen in UNETR. (d) A mixed architecture where the upper layers of the encoder and decoder are convolutional, while the deeper encoder, bottleneck, and decoder incorporate transformer blocks, as in TransUNet.*

The SETR [27] and Swin Transformer [28] have demonstrated the effectiveness of utilizing full and windowed transformer architectures for image segmentation. These methods leverage self-attention mechanisms to model long-range dependencies within feature maps, facilitating semantic information extraction. Compared to SETR, which directly adopts the fundamental transformer blocks from ViT [200], Swin Transformer introduces a shifted window operation that constrains self-attention computation to non-overlapping local windows while enabling cross-window connections. This design not only enhances computational efficiency but also achieves a well-balanced trade-off between local feature learning and global context modeling. Moreover, the hierarchical architecture of Swin Transformer enables the extraction of multi-scale features, which is crucial for handling objects of varying sizes in medical images.

In Swin-Unet [201], the fundamental Swin Transformer block is integrated into a U-Net-like encoder-decoder architecture, leveraging its hierarchical self-attention mechanism for enhanced feature representation. Experiments on multi-organ segmentation (Synapse dataset[2]) and cardiac segmentation (ACDC dataset [202]) demonstrate that this Transformer-based U-shaped network outperforms conventional approaches. These findings indicate the effectiveness of Transformer-based architectures in medical image segmentation, particularly in capturing long-range dependencies while preserving fine-grained spatial details. However, despite their promising performance, pure Transformer-based models often come with increased computational complexity and memory requirements, which can pose challenges for deployment in clinical settings or resource-constrained environments. Additionally, their performance may degrade when training data is limited, as Transformers typically require large datasets to generalize well. These limitations have led to the continued exploration of hybrid architectures that aim to balance global context modeling with computational efficiency.

---

[2] https://www.synapse.org/Synapse:syn3193805/wiki/217789

In UNETR [203] and Swin UNETR [18], Transformer blocks are incorporated into the encoder to capture global, multi-scale representations of 3D medical images, addressing the locality limitations of convolutional operations. To reduce computational overhead, convolutional blocks are employed in the bottleneck and decoder. Additionally, feature representations from various Transformer layers in the encoder are extracted and integrated into the corresponding convolutional layers of the decoder through skip connections, enhancing the final segmentation performance. Although this architecture has demonstrated strong efficacy across several benchmark datasets, the quadratic complexity of the self-attention mechanism in the encoder remains a significant computational bottleneck—particularly for high-resolution 2D images and 3D volumetric medical data.

To alleviate this limitation, hybrid architectures (Figure 32(d)) have been proposed, which restrict the application of self-attention to deeper encoder layers or selected regions of interest, thereby achieving a balance between computational efficiency and segmentation performance. For instance, TransUNet [29] introduces a Transformer-based encoder on top of a CNN backbone to preserve both global contexts and local spatial details. LeViT-UNet [30] incorporates lightweight vision Transformers to reduce computational overhead without compromising accuracy. MaskAttn-UNet [204] integrates masked self-attention to focus on informative regions, improving segmentation of small or ambiguous structures. UCTransNet [189] enhances cross-scale feature fusion by employing Transformer units at key interaction points between encoder and decoder. These hybrid approaches demonstrate that careful architectural design can harness the strengths of both CNNs and Transformers while mitigating the cost of full attention computation in high-dimensional medical imaging tasks.

## 3.6 Representative fully supervised learning methods for Medical Image Segmentation

Table 1 summarizes a selection of seminal fully supervised learning methods for medical image segmentation discussed in the preceding sections. For each method, we list the method name, publication year, key contributions, primary image modality, and corresponding source code from GitHub.

Table 1. Representative fully supervised learning–based methods for medical image segmentation.

| Method | Year | Key Contribution | Modality | Source Code |
|--------|------|------------------|----------|-------------|
| U-Net [15] | 2015 | Encoder–decoder with skip connections for precise localization; small data efficiency | Biomedical microscopy, general medical imaging | https://github.com/milesial/Pytorch-UNet |
| V-Net [16] | 2016 | 3D fully convolutional network for volumetric segmentation with Dice loss | MRI volumetric segmentation | https://github.com/faustomilletari/VNet |
| nnU-Net [17] | 2021 | Self-configuring framework that adapts preprocessing, architecture, training to dataset | Multi-modal segmentation (CT, MRI, PET) | https://github.com/MIC-DKFZ/nnUNet |
| Attention U-Net [182] | 2018 | Attention gates for focusing on relevant spatial regions | CT, MRI lesion segmentation | https://github.com/ozan-oktay/Attention-Gated-Networks |
| UNet++ [192] | 2018 | Nested and dense skip connections to reduce semantic gap between encoder and decoder | Multi-organ segmentation | https://github.com/MrGiovanni/UNetPlusPlus |
| TransUNet [29] | 2021 | Combines Transformer encoder with CNN decoder for global context | Multi-modal MRI/CT segmentation | https://github.com/Beckschen/TransUNet |
| LeViT-UNet [30] | 2021 | Lightweight hybrid CNN–Transformer architecture (LeViT backbone) integrated into a U-Net framework for fast and accurate medical image segmentation | CT, MRI | https://github.com/apple1986/LeViT-UNet |
| Swin-UNet [201] | 2022 | Hierarchical Swin Transformer for medical image segmentation | CT, MRI | https://github.com/HuCaoFighting/Swin-Unet |
| SwinUNETR | 2022 | Combines a Swin Transformer encoder with a UNETR-style decoder for high-performance 3D CT/MRI segmentation. | 3D CT, 3D MRI | https://monai.io/research/swin-unetr |
| SAM [31] | 2023 | Promptable image segmentation foundation model trained on 1B masks; adaptable to medical | Natural & medical images | https://github.com/facebookresearch/segment-anything |

| | | imaging via fine-tuning | | |
|---|---|---|---|---|
| SAM2 [35] | 2024 | Next-generation SAM with improved memory mechanisms for efficient multi-object and video segmentation; better performance in domain adaptation | Natural & medical images, video sequences | https://github.com/facebookresearch/segment-anything-2 |

## 4. Semi-supervised segmentation

Although U-Net-based fully supervised approaches have demonstrated significant progress in various medical image segmentation tasks [17], these methods typically require a substantial amount of high-quality labeled data for training. However, annotating medical images is both time-consuming and expensive, as it relies on the expertise of human annotators. Semi-supervised medical image segmentation offers an effective solution by leveraging a small number of labeled images alongside a large volume of unlabeled images to enhance model performance [88]. Broadly, semi-supervised segmentation methods can be categorized into consistency regularization methods, pseudo-labeling methods, and prior knowledge-based methods, based on their adherence to the following three key principles.

(1). Smoothness Assumption: Points that are close to each other in the feature space are more likely to share the same label. This implies that the decision boundary should pass through regions with low data density, ensuring that similar samples are classified similarly.

(2). Cluster Assumption: Data points within the same cluster (high-density regions) are likely to have the same label. This assumption is closely related to the smoothness assumption but emphasizes the clustering structure of the data, where clusters of similar data points correspond to specific classes.

(3). Manifold Assumption: The data lies on or near a lower-dimensional manifold within the high-dimensional input space. This means that the model can effectively learn from fewer labeled data points by focusing on the structure of the data's intrinsic manifold, which the unlabeled data helps reveal.

In Section 4.1, we describe three categories of consistency regularization methods: data-level, feature-level, and model-level consistency. Section 4.2 provides a brief overview of pseudo-labeling approaches for semi-supervised segmentation. Finally, Section 4.3 reviews several classical methods that incorporate prior knowledge.

## 4.1 Consistency regularization

Consistency regularization holds the smoothness assumption that predictions of unlabeled data should be invariant to different forms of perturbations [205]. Roughly, there are three types of consistency regularization approaches according to which level the perturbation happens: data-level, feature-level, and model-level consistency. The main difference among these methods is illustrated in Figure 33.

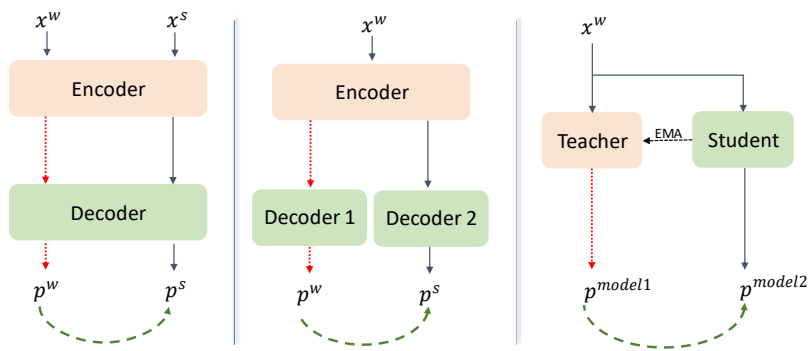

*Figure 33. Illustration of data-level consistency, feature-level consistency, and model-level consistency. Here, $x^w$ and $x^s$ represent the weakly and strongly augmented data from the same unlabeled input $x$, while the $p^w$ and $p^s$ denote the corresponding predictions. The prediction $p^w$ serves as a pseudo-label for model training with the input $x^w$. The red dashed lines indicate gradient backpropagation, and the green dashed curves represent supervision. EMA: exponential moving average.*

### (1) Data level consistency

Data-level consistency involves applying various perturbation functions to the input data while ensuring that the corresponding outputs remain consistent. This approach operates under the assumption that weakly augmented images can produce reliable predictions, while strongly augmented images enhance the learning process and improve model robustness [39]. Typically, weakly augmented images (e.g., those subjected to rotations, translations, or scaling) are employed to generate pseudo-labels from the model's predictions. In contrast, strongly augmented images (e.g., those with contrast variations, blurring, or added noise) are used to train the model using these pseudo-labels, thereby reinforcing its consistency and robustness.

FixMatch is one of the pioneering works in semi-supervised learning, leveraging weakly and strongly perturbed input data to enforce prediction consistency on the corresponding outputs with high confidence [206]. Specifically, it first generates pseudo-labels from weakly augmented inputs, retaining only those with high-confidence predictions. These retained pseudo-labels serve as target reference labels for the strongly augmented versions of the same inputs. The model is then further trained by enforcing consistency regularization between the reference pseudo-labels and the strongly augmented inputs. Adopting the principle of FixMatch, many data consistency-based methods are explored for semi-supervised segmentation using weakly augmented and strongly augmented samples, such as mixup in CCAT-NET [207], copy-paste in BCP [208], and elastic deformations used in a Siamese architecture [209], etc. These methods employ different data transformation strategies for input images, aiming to ensure consistent segmentation results across the paired weakly augmented and strongly augmented data.

While data-level consistency effectively leverages the rich information contained within the data, it does not guarantee the generation of high-quality pseudo-labels from weakly augmented images. Poor-quality pseudo-labels can violate the consistency assumption and reduce the model's accuracy. For instance, small perturbations in weakly augmented images may introduce noise or outliers, leading the model to focus on these noisy regions and produce low-quality pseudo-labels. Therefore, the choice of augmentation strategies for both weakly and strongly augmented images must be carefully considered to maintain label consistency during training [210]. In addition, not all augmentation strategies commonly used for natural images are clinically meaningful or beneficial in medical image segmentation. Certain perturbations can contradict anatomical reality and mislead the model. For example, flipping along the axial or coronal planes is generally meaningless, and even sagittal flipping is often inappropriate, as human anatomy is rarely perfectly symmetric. Similarly, excessive contrast variation, blurring, or artificial noise may distort essential clinical features and reduce the validity of the learning process. Thus, augmentation design in medical imaging must carefully account for anatomical plausibility and modality-specific constraints.

## (2) Feature level consistency

In addition to perturbations at the image level, many studies have explored feature-level disturbances to maintain consistency, guided by specific consistency-based objective functions. In CCT [211], perturbations are applied to the encoder's output features and subsequently processed by multiple decoders. This method seeks to learn invariant predictions for the same input image, based on the premise that the decision boundary is more discernible at the hidden feature level than at the image level. FeatMatch [212] introduces a learned feature-based refinement and augmentation method designed to generate a diverse range of feature transformations. These transformed features are incorporated into the consistency-based regularization loss, enhancing the model's robustness. Moreover, virtual adversarial training (VAT) [213] brings a novel approach by introducing adversarial feature perturbations. VAT has been integrated into two teacher models and applied to a student model in PS-MT [214], with the goal of improving the model's resilience to challenging noise and enhancing its generalization performance. In UCD-Net [215], seven types of feature perturbations, including feature noise, feature dropout, object masking, context masking, guided cutout, intermediate VAT and random dropout, are utilized in seven auxiliary decoders, seeking for these predictions from decoders to be consistent with the main decoder (see Figure 34).

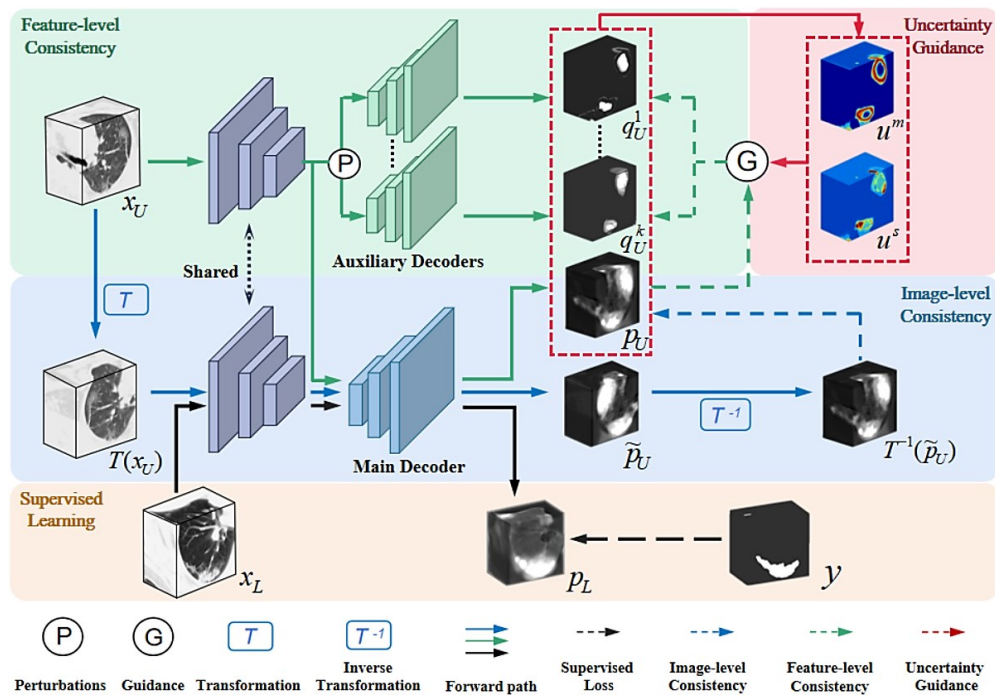

*Figure 34. Structure of UDC-Net. Feature-level consistency (in green) is leveraged between the main decoder's prediction and seven other auxiliary decoders' predictions. From [215].*

Feature-level consistency, like image-level consistency, is rooted in the cluster assumption for unlabeled data. Consequently, many approaches integrate perturbations at both the image and feature levels to introduce additional information and mitigate confirmation bias, thereby improving the effectiveness of training models with unlabeled data [205, 216, 217]. However, significant challenges remain, particularly in determining the quality of pseudo-labels and selecting optimal perturbation methods for medical image semi-supervised segmentation. To address these issues, some studies have introduced strategies such as adaptive thresholds [218], self-ensembling [219], uncertainty-aware [220], discriminator-based classifiers [221], pairwise relation learning [222], and so on.

### (3) Model level consistency

Unlike perturbations on images or feature maps, model-level consistency refers to the principle that different models or different versions of the same model should produce consistent predictions when presented with the same input data, regardless of variations in their architecture, training conditions, or initialization. This approach leverages the idea of ensemble learning, temporal ensembling, or teacher-student frameworks to improve the robustness and generalization of the model by aligning their outputs [223-225].

One of the representative model-level consistency frameworks is the Mean Teacher (**MT**) model [223] (see Figure 35). In this framework, the teacher model's weights are updated as an exponential moving average (EMA) of the student model's weights. The teacher model generates pseudo-labels for unlabeled data, which are then used to train the student model. Consistency is enforced by aligning the student's predictions with those of the teacher. Building on the success of the MT framework for semi-supervised image classification, an adapted MT model was introduced for brain lesion segmentation in [226] and left atrium segmentation [227]. This approach leverages both labeled and unlabeled data to enhance segmentation performance. Specifically, two identical DeepMedic architectures [228] were employed as the teacher and student models, respectively.

However, a key limitation of the MT model lies in the tight coupling of weights between the teacher and student models, as the teacher is essentially an EMA of the student. This dependency can hinder the generation of high-quality pseudo-labels if either model fails to train effectively [210]. To address this limitation, the Dual Student model was proposed in [229] (See Figure 35). This approach replaces the teacher model in the MT framework with an independently initialized second student model, introducing a bidirectional stabilization constraint between the two students to maintain loosely coupled weight relationships and promote balanced mutual learning. Stable samples, which are less affected by small perturbations and are far from the decision boundary, are selected to facilitate the exchange of only reliable knowledge between the two student models. This strategy ensures stability and mitigates the variability in outputs from independently trained models. Similar to the Dual-Student framework, Cross Pseudo Supervision (CPS) enforces consistency between two segmentation networks initialized differently but processing the same input image. Each network's output serves as a pseudo-label to supervise the other network via the standard cross-entropy loss, applied in both directions [230]. Building on this concept, PS-MT [214], two teacher models with a single student model are proposed to mitigate inaccurate predictions from the teacher models.

In addition, additional data-level transformations and feature-level perturbations can be integrated into the teacher-student framework, aiming to improve the quality of pseudo-labels generated by the teacher models [205, 231-234]. Alternative designs include architectures that share a common encoder but employ different decoders, as demonstrated in MC-Net [235], URPC [236], and DTC [237].

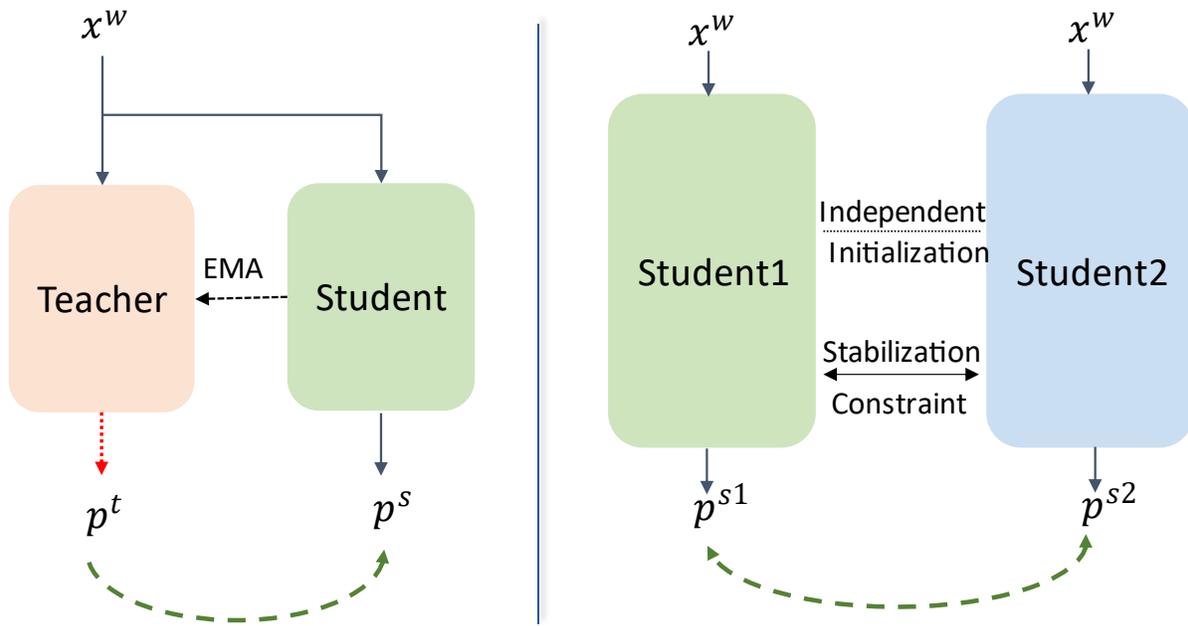

*Figure 35 Two types of model-level consistency frameworks. Left: The dual-model framework consisting of a teacher–student pair, exemplified by the Mean Teacher approach, in which the teacher's weights are tightly coupled to the student's via exponential moving average (EMA). Right: The dual-student framework, which employs two independently initialized student models with loosely coupled weights, enforcing mutual consistency through a bidirectional stabilization constraint.*

## 4.2 Pseudo-labeling

Pseudo-labeling is a technique in semi-supervised learning that leverages unlabeled data by generating artificial labels, typically using a model's own predictions. The process of semi-supervised semantic segmentation using pseudo-labeling, as illustrated in Figure 36, generally follows these steps: (1). the model is trained using labeled data; (2). The trained model is used to generate pseudo labels for unlabeled images; (3). The labeled and pseudo-labeled data are combined to retrain the model; and (4). this retrained model is subsequently used to update the pseudo labels on the unlabeled data and labeled data. Steps (2) through (4) are repeated iteratively until the model's performance saturates. Typically, two major paradigms in pseudo-labeling for semi-supervised segmentation are self-training and co-training. The key difference between the two lies in the way pseudo-labels are used: in co-training, pseudo-labels generated from one view are added to the training set and used to supervise models trained on other views. The self-training approach differs by iteratively refining pseudo-labels generated by a single model.

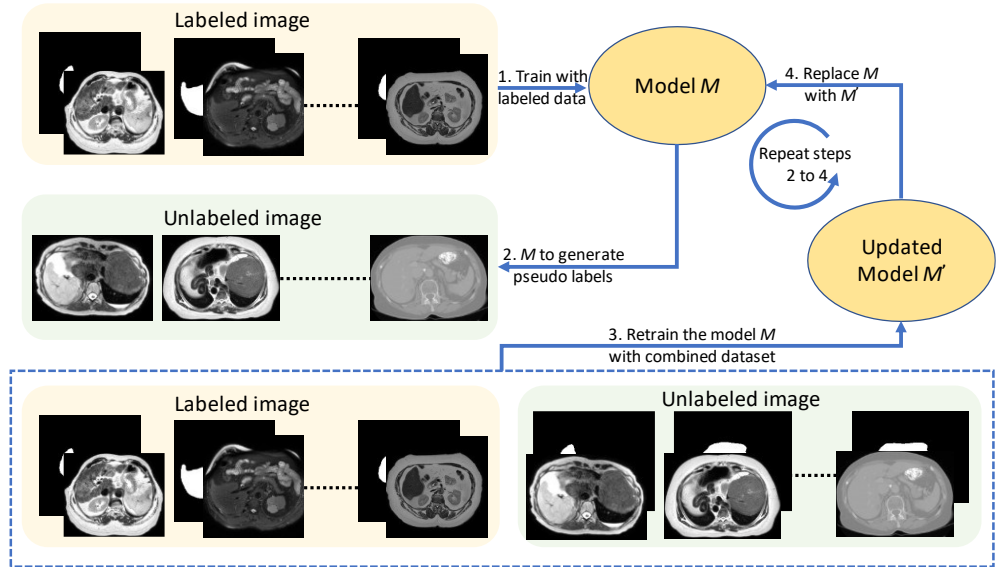

*Figure 36. The pipeline of pseudo-labeling for semi-supervised segmentation.*

## (1) Self-training

Self-training methods are the most basic of pseudo-labeling approaches. It is a widely used semi-supervised learning strategy where a model is initially trained on a small set of labeled data, and then used to generate pseudo-labels for unlabeled data [238]. These pseudo-labeled samples, often selected based on the model's confidence, are then incorporated into the training set to refine the model further. Due to its simplicity and ease of implementation, self-training has become a foundational technique in semi-supervised learning. Subsequent research has focused on enhancing various aspects of this framework, including confidence thresholding [239], sample selection strategies [240], consistency regularization [241], and dynamic reweighting of pseudo-labeled samples [242] to improve robustness and reduce error accumulation.

In medical image segmentation, self-training has gained popularity due to the high cost and specialized expertise required for manual annotation. By leveraging large volumes of unlabeled imaging data, self-training approaches have shown promise in improving segmentation accuracy, particularly in challenging scenarios such as rare disease detection, small lesion segmentation, or cross-modality adaptation, where labeled examples are scarce. For example, Bai et al. [243] demonstrated that self-training significantly improved cardiac segmentation performance with only a small set of labeled data. Similarly, in [244], pseudo-labels were generated within the Inf-Net framework using a randomly selected propagation strategy to mitigate the lack of annotated samples. In [219], a transformation-consistent self-ensembling approach was proposed for 3D CT segmentation, which reinforced pixel-level prediction consistency through data augmentation. Improving the quality of pseudo-labels remains central to the success of self-training algorithms. To this end, various strategies have been explored, including shape prior-based selection [245], uncertainty bootstrap integration [246], threshold selection [247], uncertainty estimation [248] etc. Although self-training algorithms are simple and easy to implement, their effectiveness can be negatively impacted when the initial pseudo-labels contain significant noise. Noisy labels may mislead the training process and degrade model performance. Therefore, improving the quality of pseudo-labels and identifying optimal strategies for selecting reliable pseudo-labeled samples for retraining remain active areas of research [249, 250].

## (2) Co-training

Co-training assumes that each data sample has multiple complementary and sufficient views. Two or more models are trained on different views and iteratively label unlabeled data for each other, reinforcing the learning process. This multi-view strategy reduces reliance on labeled data and enhances generalization. Co-training, by independently updating all models via backpropagation, leverages view diversity to improve robustness [39, 251].

The core of co-training lies in constructing different views and maintaining consistent predictions across all views. Broadly, there are three basic strategies to create different views: (i) applying different input transformations [251, 252], (ii) leveraging different image modalities [253, 254], and (iii) using different model architectures [255, 256]. Often, these strategies are combined within the co-training framework.

For instance, a method called *uncertainty-aware multi-view co-training* was proposed in [251]. The authors first apply input transformations, such as rotations and permutations of 3D volumes, to create multiple views (e.g., axial and coronal views), and then train a separate 3D neural network for each view. Co-training is applied by enforcing multi-view consistency on unlabeled data, with the aim of improving performance across all views. In [253], unpaired CT and MRI images are leveraged to extract modality-independent knowledge for semi-supervised segmentation. Specifically, cross-modal consistency between CT and MRI is employed as a supervision signal for unlabeled data. To effectively incorporate this knowledge into the segmentation model, the authors propose an atlas-based framework consisting of two modality-specific encoders and a shared decoder. This architecture is designed to learn generalized and robust modality-independent features by enforcing both semantic and anatomical consistency across modalities. In [255], a self-correcting co-training scheme is proposed to enhance the quality of pseudo-labels based on collaborative network outputs. Specifically, a self-correcting module is introduced to improve the learning confidence for unlabeled data. Additionally, a structural constraint is imposed to regularize the shape similarity between the current predictions of the collaborative networks and the final improved learning target (pseudo-label) (see Figure 37).

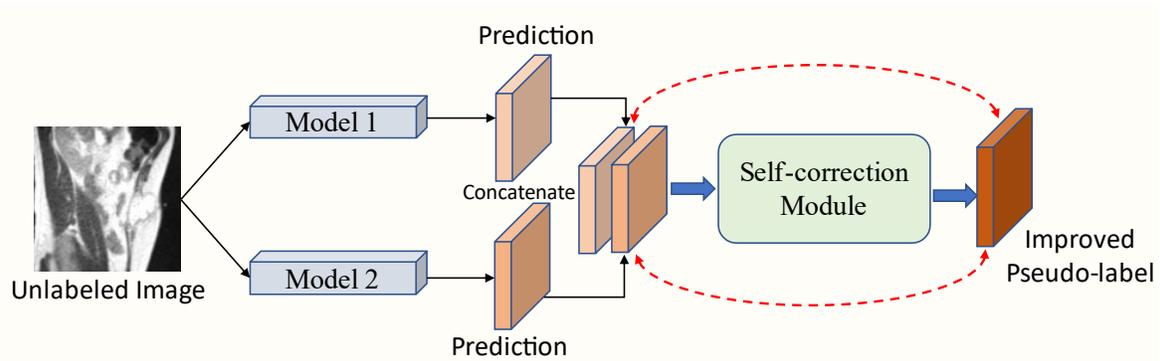

*Figure 37. Overview of the self-correcting pipeline used to enhance pseudo-label quality for unlabeled data in semi-supervised segmentation.*

Co-training, while effective, presents several notable limitations. First, it assumes that each view from the input is independent when making predictions; however, this assumption may not hold in clinical applications. In cases where views are correlated or redundant, this assumption can lead to suboptimal performance. Second, co-training strives to maintain consistency between the predictions of multiple models using different architectures. However, if these models are overly similar, they may fail to extract sufficiently discriminative representations from the data, thereby limiting the approach's effectiveness. Third, co-training relies on pseudo-labels generated from one view, which are subsequently incorporated into the training set. If these pseudo-labels are erroneous, their inclusion can negatively impact model performance. Thus, ensuring the generation of high-quality pseudo-labels remains a fundamental challenge in co-training. A promising avenue to address this issue is the utilization of foundation models, such as the Segment Anything Model (SAM) [31, 35], to generate reliable pseudo-labels. Techniques like SAMatch [257] and SemiSAM [258] offer potential solutions for improving the accuracy and quality of pseudo-labels in co-training frameworks.

## 4.3 Prior knowledge-based semi-supervised segmentation

Medical images inherently contain rich prior knowledge, such as anatomical shape, spatial position, relative scale, and key landmarks (see Figure 38). Effectively incorporating this prior knowledge into segmentation methods can not only enhance segmentation accuracy but also improve the efficiency of object recognition [240] and object delineation [2], accelerating both training and inference processes.

Currently, two main categories of prior knowledge are widely incorporated into semi-supervised segmentation to boost performance: (i) structural priors, and (ii) statistical or learning-based priors.

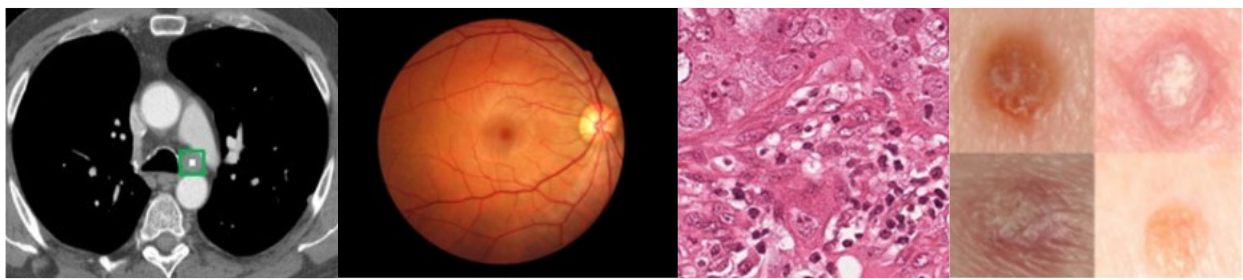

*Figure 38. Examples of four imaging modalities (from left to right: CT, fundus, pathology, and dermoscopy), each containing prior knowledge such as location, shape, intensity distribution, and texture.*

## (1) Structural priors

Structural priors primarily encompass anatomical knowledge, including the shapes, sizes, locations, and spatial relationships of objects. They also incorporate appearance-related information, such as intensity distributions and textural patterns characteristic of different tissues. In STFA [88], shape information is extracted from the structure tensor and integrated into a loss function, aiming to enforce consistency in the outputs of the same image under different augmentations. This approach demonstrates that shape priors provide greater stability and efficiency during semi-supervised segmentation training. In BaPC [259], boundary-aware prototypes are employed to explicitly model the differences between the boundaries and centers of adjacent object classes, thereby enhancing discriminative feature learning. In this context, a prototype is regarded as the most representative feature of a category in the feature space, effectively bridging the gap between labeled and unlabeled data for semi-supervised segmentation.

In addition to directly extracting structural priors, atlas-based approaches are widely used to integrate positional information and shape priors into semi-supervised segmentation frameworks. For example, DeepAtlas [260] incorporates image registration into a segmentation framework to fully exploit unlabeled images during training. An atlas is first constructed from the training images using affine registration. Subsequently, the registration and segmentation networks are jointly optimized, interactively guiding each other's training through an anatomy similarity loss. Specifically, the segmentation sub-network provides anatomical masks for weak supervision of the registration network, while the registration network improves the segmentation sub-network's learning on unlabeled data. This mutual supervision strategy enables more effective use of limited annotations and improves performance in both tasks.

In [261], a probabilistic atlas incorporating anatomical prior knowledge is introduced as a loss function, termed the Deep Atlas Prior Loss. This loss is integrated into an adversarial learning-based semi-supervised segmentation framework [262], serving as accurate organ segmentation during training. Experimental results on liver and spleen datasets demonstrate that the proposed loss function outperforms traditional loss functions such as cross-entropy, focal loss, and Dice loss, highlighting the effectiveness of leveraging anatomical priors to enhance segmentation accuracy. Similarly, in [263], an atlas-based semi-supervised segmentation network is proposed, where a probabilistic atlas is employed to generate pixel-wise confidence maps, enabling the generation of reliable pixel-level pseudo-labels for unlabeled data.

Although integrating structural priors into semi-supervised learning has proven effective—providing more stable and robust features to better leverage unlabeled data—there remain notable challenges. These include accurately modeling anatomical variability across populations, avoiding bias introduced by overly rigid priors, and ensuring generalizability across imaging modalities and clinical scenarios. Future research may benefit from adaptive or learnable structural priors that can dynamically adjust to the context of individual cases, thereby enhancing flexibility and robustness in semi-supervised segmentation.

### (2) Statistical and distributional priors

Statistical and distributional priors encode knowledge about the underlying data distributions, label distributions, and sampling biases inherent in medical imaging datasets. In semi-supervised settings, these priors are commonly leveraged to align feature distributions between labeled and unlabeled data, typically through domain adaptation techniques such as distribution-matching losses based on maximum mean discrepancy [264], adversarial learning [265], and entropy minimization [266].

Building on this idea, several recent studies have further explored how to explicitly model and exploit distribution differences for more robust semi-supervised segmentation. For example, in [267], statistical differences across datasets are explicitly leveraged to improve generalizable semi-supervised medical image segmentation, where a limited amount of labeled data is combined with abundant unlabeled data from diverse cohorts. To address domain shifts, the framework introduces multiple statistic-specific branches that generate more reliable pseudo-labels tailored to individual domains. In parallel, a statistics-aggregated branch is employed to learn domain-invariant features, promoting consistent performance across varying data distributions. In [268], a method incorporating distribution calibration and a non-local semantic constraint is proposed to address distribution misalignment between labeled and unlabeled data in semi-supervised segmentation. Specifically, hidden features extracted from both labeled and unlabeled samples are recalibrated by minimizing the L2 distance between their distributions, guided by a non-local semantic constraint derived from the labeled data. Together, these approaches highlight the importance of addressing statistical misalignment to fully unlock the potential of unlabeled data in semi-supervised segmentation.

## (3) Functional and clinical priors

Functional priors capture continuous dynamic processes in both time and anatomical structure, such as temporal sequences in dynamic imaging modalities (e.g., 4D MRI, cine MRI), disease progression patterns, and physiological phenomena like blood flow [269]. In this context, [270] introduces translation consistency as a structural prior for 3D semi-supervised medical image segmentation. Unlike conventional perturbation-based consistency learning methods—which typically preserve the spatial context across different views and may cause models to overfit to background structures—translation consistency explicitly disrupts spatial alignment while preserving object integrity. This encourages the model to learn segmentation patterns that are truly object-centric, rather than context-dependent, thereby improving robustness in scenarios with limited annotations. Furthermore, disease-specific priors, such as tumor motion models in radiotherapy or cancer staging systems, provide valuable context for semi-supervised learning by constraining predictions to clinically plausible patterns. These priors can be embedded into the model through rule-based constraints, hierarchical loss functions, or graph-based label propagation.

Functional and clinical priors guide models to produce outputs that are not only mathematically optimal but also physiologically and clinically meaningful, which is critical for real-world adoption of AI systems in healthcare.

## 4.4 Seminal semi-supervised learning methods for medical image segmentation

Table 2 presents a selection of representative semi-supervised learning methods for medical image segmentation. For each method, we provide the method name, year of publication, key contributions, primary application domain, and corresponding GitHub source code.

| Method | Year | Key Contribution | Application Domain | Source Code |
|--------|------|-----------------|-------------------|-------------|
| UA-MT [227] | 2019 | Combines mean teacher consistency with uncertainty weighting | 3D medical image segmentation | https://github.com/yulequan/UA-MT |
| FixMatch [206] | 2020 | Combines consistency regularization with pseudo-labeling using strong/weak augmentations | Multi-domain segmentation | https://github.com/kekmodel/FixMatch-pytorch |
| CPS [230] | 2021 | Two-network pseudo label exchange to improve consistency | General image segmentation; adapted to medical | https://github.com/charlesCXK/TorchSemiSeg |
| DTC [237] | 2021 | Enforces consistency between segmentation and boundary prediction | Multi-organ CT segmentation | https://github.com/HiLab-git/DTC |
| MC-Net [235] | 2021 | Multi-branch co-training with temporal ensembling | Cardiac MRI segmentation | https://github.com/ycwu1997/MC-Net |
| URPC [236] | 2021 | Uncertainty Rectified Pyramid Consistency for semi-supervised learning | CT organ segmentation | https://github.com/HiLab-git/URPC |
| UniMatch [205] | 2022 | Unified semi-supervised segmentation framework that balances strong and weak consistency constraints with adaptive thresholding | General image segmentation; adapted to medical | https://github.com/LiheYoung/UniMatch |
| UniMatch v2 [231] | 2023 | Improves UniMatch with multi-view consistency, dynamic confidence thresholds, and stronger augmentation strategies | General image segmentation; adapted to medical | https://github.com/LiheYoung/UniMatch |
| SAMatch [257] | 2024 | Leverages the Segment Anything Model (SAM) to generate high quality pseudo-labels for U-Net | Multi-organ and lesion segmentation in CT/MRI/US | https://github.com/apple1986/SAMatch |

## 5. Discussion

As medical image segmentation continues to evolve, several critical shifts are accelerating the transition from fully supervised learning toward semi-/weakly supervised and unsupervised paradigms. These shifts are largely driven by practical challenges, including data scarcity, domain variability, and the demand for scalable, generalizable solutions. In this section, we examine seven key themes shaping the future trajectory of the field: label-efficient learning, lesion-focused segmentation, multi-modality domain adaptation, transfer learning with foundation models, probabilistic segmentation, and advanced 3D/4D segmentation methodologies.

## 5.1 From fully supervised to semi-supervised and unsupervised segmentation

With the advancement of medical image segmentation, two prominent trends have emerged. The first focuses on the collection and creation of large-scale public datasets to facilitate standardized and comparative evaluation of segmentation algorithms. Notable examples include the Medical Segmentation Decathlon (MSD) [271], AbdomenAtlas-8K [272], AMOS [273], BTCV [274], BraTS [275] and MedSAM2 [276] [3]. These efforts have primarily supported the development of fully supervised methods, including foundation models. The second trend emphasizes leveraging limited labeled data in conjunction with abundant unlabeled data within a semi-supervised learning framework. In many practical scenarios, where high-quality or pixel-level annotations are scarce, segmentation frameworks are extended to weakly supervised settings—using coarse, inexact, or indirect labels—or to fully unsupervised approaches that rely solely on the intrinsic structure of the data.

In fully supervised segmentation, a central challenge lies in how to *fully leverage* annotated data to learn robust and discriminative features under diverse conditions, such as varying modalities and image noise. With the emergence of large-scale foundation models for medical image segmentation, it may appear that the field is approaching its "holy grail." However, several critical challenges remain unresolved. These include: (1) improving efficiency with respect to time, computational resources, and inference speed; (2) addressing class imbalance in large-scale datasets, particularly for underrepresented or hard-to-segment structures with limited annotations; and (3) uncovering theoretical insights and interpretability mechanisms to better understand how deep neural networks perform segmentation in medical imaging.

For semi-supervised learning, the primary motivation is to generate high-quality pseudo-labels for unlabeled data through novel algorithmic strategies. Although recent methods have achieved performance on par with fully supervised models on benchmark datasets such as ACDC [257] and LA [258], substantial challenges remain when transitioning to clinically relevant datasets sourced from multiple institutions and heterogeneous imaging devices. Several open problems could be considered to fully capitalize on unlabeled data for performance enhancement: (1) how to effectively incorporate prior knowledge into semi-supervised frameworks—for instance, via cross-modal feature fusion from images, language, and handcrafted features; (2) how to manage distributional mismatch between labeled and unlabeled datasets, which can violate the assumption of data consistency and potentially degrade model performance [277]; and (3) how to generate and utilize optimal pseudo-labels. Some approaches, such as UniMatch [205], employ confidence thresholding on probability maps to derive pseudo-labels; however, this strategy may discard critical structural information (e.g., object shape and connectivity) or propagate erroneous pixel-level labels.

---

[3] This website curates accessible medical image segmentation datasets: https://medsam-datasetlist.github.io/

The weakly supervised and unsupervised segmentation paradigms may provide complementary solutions in scenarios where precise annotations are unavailable or prohibitively expensive to obtain. Weakly supervised approaches leverage coarse annotations such as bounding boxes, scribbles, or image-level labels to approximate segmentation masks, thereby reducing the annotation burden while still guiding the learning process. Unsupervised methods, on the other hand, rely entirely on the inherent structure and distribution of the data—using techniques such as clustering [278], self-supervised learning [279], or contrastive representation learning [280, 281]—to discover semantic boundaries without any human-provided labels. Despite their promise, both paradigms still face critical limitations. Weakly supervised methods must contend with the challenge of learning accurate spatial representations from inherently ambiguous or imprecise labels. Unsupervised approaches often struggle to align discovered clusters with meaningful anatomical structures, especially in complex medical images with high inter-patient variability and subtle pathological features. Future research may benefit from hybrid strategies that combine weak and self-supervised signals or incorporate domain knowledge and anatomical priors to guide the learning process more effectively. While we fully acknowledge the importance of establishing strong theoretical underpinnings for segmentation methods, it is equally important to recognize that, in clinical practice, the ultimate measure of success is the reduction in human effort required to produce segmentations of sufficient quality for the intended application. Thus, balancing methodological rigor with practical efficiency remains a critical consideration for advancing medical image segmentation toward real-world adoption.

## 5.2 From organ to lesion segmentation[4]

While both supervised and semi-supervised methods have achieved notable success in segmenting large, anatomically consistent organs, lesion segmentation—particularly for small, irregular, or poorly defined regions—remains a significant challenge. This difficulty arises primarily from high inter-observer variability, diverse lesion morphology, and the limited availability of precise annotations (see Table 3).

Table 3. Summary of representative public datasets for lesion segmentation on medical image.

---

[4] Small structures (e.g., adrenal glands), spatially sparse organs (e.g., esophagus, pancreas), highly deformable tissues (e.g., small and large bowel), and boundary-less regions (e.g., lymph node zones) remain challenging to segment. Lesion segmentation algorithms should ideally be capable of addressing these cases, despite differences in characteristics.

| Dataset | Year | Scans (Train/Test) | Lesion/Target | Region | Modality | Description |
|---|---|---|---|---|---|---|
| BraTS [275] | 2012 | 50/15 | Brain tumors (gliomas) | Brain | MRI | Multi-parametric MRI (T1, T1Gd, T2, FLAIR) with expert annotations for tumor subregions, including enhancing tumor, tumor core, and whole tumor. Released annually with varying train/test splits: BraTS 2018 (210/75) [282], BraTS 2019 (256/76) [283], BraTS 2020 (369/125) [284], BraTS 2021 (~2,000 cases) [285], and BraTS 2023 (~2,400 cases) [286] |
| ISLES [287] | 2015 | 28/36 (SISS); 30/20(SPES) | Ischemic stroke lesions | Brain | MRI | Multi-sequence MRI (DWI, FLAIR, T2) for acute and subacute stroke lesion segmentation, comprising two subsets: Sub-Acute Stroke Lesion Segmentation (SISS) and Stroke Perfusion Estimation (SPES); later expanded to ISLES16/17[288]and ISLES24[289] |
| ATLAS [290] | 2018 | 955/245 | Stroke lesions | Brain | MRI | Anatomical Tracings of Lesions After Stroke. T1-weighted MRI with manually delineated lesions. |
| SegTHOR [291] | 2020 | 40 / 20 | Thoracic tumors (esophageal) | Thorax | CT | Segmentation of esophagus, heart, trachea, and tumor from thoracic CT scans. |
| HECKTOR [292] | 2020 | 201 / 53 | Head & neck squamous cell carcinoma | Head & Neck | PET/CT | Multi-institutional dataset for tumor segmentation in head and neck cancer; expanded to 224/101 cases in 2021 [293] and 524/359 cases in 2022 [294]. |
| LiTS [295] | 2017 | 131 / 70 | Liver tumors | Abdomen | CT | Contrast-enhanced abdominal CT scans for liver and lesion segmentation. |
| KiTS19 [296] | 2019 | 210 / 90 | Kidney tumors | Abdomen | CT | Kidney and kidney tumor segmentation from contrast-enhanced abdominal CT scans; expanded in KiTS21 to 300/100 cases in 2021. |
| MSD – Hepatic Vessel [271, 297] | 2019 | 303 / 140 | Liver tumors & vessels | Abdomen | CT | From Medical Segmentation Decathlon (Task 08). Contrast-enhanced abdominal CT with hepatic vessel and tumor labels. |
| MSD – Pancreas [271, 297] | 2019 | 282 / 138 | Pancreatic tumors | Abdomen | CT | From Medical Segmentation Decathlon (Task 07). Abdominal CT for pancreas and pancreatic tumor segmentation. |
| MSD – Lung [271, 297] | 2019 | 63 / 33 | Lung tumors | Abdomen | CT | From Medical Segmentation Decathlon (Task 06). Thoracic CT for lung cancer segmentation. |
| MSD – Colon [271, 297] | 2019 | 126 / 64 | Colon tumors | Abdomen | MRI | From Medical Segmentation Decathlon (Task 10). Pelvic MRI for colorectal cancer segmentation. |
| PROMISE12 [298] | 2012 | 50/50 | Prostate lesions | Pelvis | MRI | T2-weighted prostate MRI for prostate glan |

Future research in lesion segmentation may benefit from several promising directions: (1) Leveraging anatomical priors—such as organ boundaries or typical lesion occurrence patterns—may enhance localization and guide feature extraction specific to certain lesion types. This organ-aware modeling can help disambiguate lesions that share visual similarities with surrounding tissue or occur in low-contrast regions; (2) The scarcity of large-scale, diverse lesion datasets—especially those covering whole-body or multi-modality imaging—motivates the use of generative models to synthesize realistic tumors. Prior studies [299-302] have demonstrated the feasibility of generating synthetic lesions in CT for specific organs such as the liver, pancreas, and kidneys. However, scaling this strategy to broader anatomical coverage and varying modalities remains an open area for further investigation. (3) In time-critical clinical scenarios, such as radiation oncology, real-time lesion segmentation is essential. Achieving this goal demands models that balance inference speed with high accuracy. Future work should focus on integrating fast segmentation frameworks with domain-specific priors—such as lesion shape or spatial context—to enable clinically viable, efficient deployment. (4)

## 5.3 From single- to multi-modality domain adaptation

Domain adaptation has been widely explored in single-modality settings, where the objective is to transfer knowledge from a labeled source domain (e.g., T1-weighted MRI) to an unlabeled or sparsely labeled target domain of the same or similar modality (e.g., T2-weighted MRI). However, real-world clinical scenarios often involve multi-modality imaging—such as MRI, CT, PET, and ultrasound—each capturing complementary anatomical and functional information. As a result, the field is increasingly shifting from single-modality to multi-modality domain adaptation (MMDA), aiming to exploit shared semantics across modalities while bridging modality-specific differences.

Compared to single-modality domain adaptation, MMDA introduces greater complexity and coordination challenges. Study [303] emphasized inter-modality consistency for fine-grained action recognition (e.g., subtle motions in surgical scenarios) but risked redundancy during fusion. The "Split to Merge" strategy in [304] aligned features pre-fusion to address heterogeneity yet struggled with scalability for high-dimensional data. In 3D segmentation, [305] leveraged mutual information between 2D and 3D data, though modality asymmetry hindered stable domain adaptation.

To advance MMDA, diverse methods have been proposed, including adversarial alignment [306], matrix matching [307], pseudo-modality generation [308], attention-guide multi-granularity fusion [309] and spatial transformations [310]. While each approach offers strengths, they face common limitations such as instability in training, reliance on strong modality correlations, vulnerability to noisy or missing data, and difficulty bridging semantic gaps across modalities. Despite technical progress, existing methods often lack robustness, efficiency, and generalizability in real-world settings with variable data quality and modality alignment.

Hence, the future research directions can be included as follows: (1) Incorporating self-supervised learning to pre-train modality-aware encoders that can be fine-tuned with limited labels; (2) Leveraging multi-modal pre-trained foundation models (e.g., MedCLIP [311] and MedCLIP-SAM [312]) that encode cross-modal correlations at scale. (3) Exploring causal domain adaptation to model modality shifts through structural relationships rather than purely statistical alignment.

## 5.4 From task-specific models to prompt-based foundation models via transfer learning

Traditional task-specific segmentation models are typically trained on narrow, predefined datasets and target limited object categories. In contrast, the advent of large-scale foundation models—pretrained on diverse and extensive datasets, such as [313], DeepSeek-V3 [314], SAM [31] and SAM2 [35]—have demonstrated impressive generalization capabilities across tasks and modalities, enabling more effective transfer learning for downstream segmentation applications. Despite these advances, fine-tuning such large models remains challenging. For example, $\pi$-Tuning interpolates between expert models to facilitate multi-task adaptation, yet errors in task similarity estimation can hinder overall performance [315]. LoRA (Low-Rank Adaptation) reduces computational overhead by decomposing weight updates into low-rank matrices, but performs sub-optimally when dealing with high-rank parameters commonly found in vision models [316]. QLoRA integrates quantization with LoRA to further reduce model size and memory usage, though the dequantization step can introduce latency during inference, especially on resource-constrained devices [317]. In addition, Adapters [318] introduces small bottleneck networks into the transformer blocks of the image encoder for finetuning image segmentation models. Representative examples include the Medical SAM Adapter (Med-SA) [319], SAM-Adapter [320] and DD-Adapter [321].

While foundation models offer promising flexibility and scalability, several key challenges remain in adapting prompt-based foundation models to medical image segmentation. Future research directions include: (1) how to effectively integrate domain-specific priors (e.g., anatomical knowledge or imaging protocols) into prompt design to improve segmentation accuracy; (2) methods to enhance robustness in scenarios with limited labeled data or missing modalities, which are common in clinical settings; (3) strategies for optimizing computational efficiency during fine-tuning and inference, ensuring feasibility for deployment in real-time or edge environments such as intraoperative imaging or point-of-care diagnostics. Furthermore, foundation models may hold potential for accelerating ground-truth annotation in both fully supervised and semi-supervised settings; however, this benefit remains to be rigorously validated through empirical studies; and (4) standardized prompt protocols to ensure cross-site reproducibility to solve prompt sensitivity and inconsistent human inputs.

## 5.5 From deterministic to probabilistic segmentation

Traditional approaches to medical image segmentation typically produce deterministic outputs, yielding a single segmentation mask for the target structure. However, in clinical practice, medical images often contain low-contrast boundaries and motion artifacts, resulting in multiple plausible segmentations for the same anatomical target. Deterministic outputs fail to capture this inherent ambiguity, providing a single mask without indicating areas of boundary uncertainty or the existence of equally reasonable alternative segmentations. This limitation can lead to overconfidence in potentially unreliable outputs, posing risks in clinical decision-making. By explicitly modeling uncertainty, segmentation methods can more accurately reflect the variability present in medical imaging data and highlight regions that may require clinician review. This growing emphasis on uncertainty has driven the integration of probabilistic models and uncertainty quantification techniques into modern segmentation pipelines, enhancing transparency, trust, and safety in clinical deployment.

In medical image segmentation, uncertainty quantification enhances clinical reliability by addressing both aleatoric and epistemic uncertainties. Aleatoric uncertainty, also known as data uncertainty, arises from inherent noise, imaging artifacts, and inter-patient anatomical variability, reflecting irreducible ambiguities in modalities such as PET and US. In contrast, epistemic uncertainty, or model uncertainty, results from limitations in the model's knowledge due to insufficient training data, suboptimal model architectures, or the presence of out-of-distribution inputs. Probabilistic segmentation methods incorporate heteroscedastic loss functions and test-time data augmentation to quantify aleatoric uncertainty, enabling the prediction of per-pixel variance alongside segmentation outputs and facilitating the generation of confidence maps that delineate regions of low reliability [322, 323]. To capture epistemic uncertainty, these methods employ Bayesian deep learning, Monte Carlo dropout, and ensemble techniques. For instance, Bayesian neural networks model epistemic uncertainty by placing distributions over network weights, while Monte Carlo dropout approximates uncertainty by sampling multiple predictions at inference [324-326].

The integration of probabilistic frameworks into semi-supervised and unsupervised learning has proven particularly transformative. In semi-supervised learning, uncertainty estimates enable the selection of high-confidence pseudo-labels for unlabeled data, thereby enhancing training stability and reducing the risk of error propagation. For instance, uncertainty-guided co-training frameworks iteratively refine pseudo-labels by leveraging confidence maps to improve label reliability. In unsupervised learning, probabilistic models facilitate anomaly detection by identifying regions of high uncertainty, such as lesions or rare pathologies that are underrepresented in the training data. Additionally, recent studies have incorporated uncertainty quantification into foundation models, further extending the capabilities of this paradigm.

A critical requirement for integrating probabilistic segmentation into clinical decision-making is that the estimated uncertainties accurately correlate with segmentation errors, ensuring that predicted uncertainty scores reliably reflect the true likelihood of correct segmentations. This alignment is achieved through uncertainty calibration [327, 328]. For instance, if a model predicts an 80% probability that a voxel belongs to a tumor, well-calibrated uncertainty implies that, across all such predictions, the voxel is indeed part of a tumor in approximately 80% of cases. To enable this, various post-hoc and integrated calibration methods have been developed. Temperature scaling is a widely adopted post-hoc technique that adjusts SoftMax outputs using a temperature parameter optimized on a validation set to align predicted probabilities with empirical frequencies. Other post-hoc methods, such as Platt scaling, isotonic regression, and histogram binning, have also been explored for calibration, although their application in segmentation tasks is less prevalent due to scalability limitations.

More recently, Dirichlet calibration extends traditional calibration methods by modeling predictive probabilities as samples from a Dirichlet distribution rather than as point estimates, enabling the capture of uncertainty across multiple classes simultaneously [329]. This approach has demonstrated superior calibration performance compared to temperature scaling in multi-class segmentation tasks. Evidential deep learning further integrates uncertainty estimation and calibration into the training process by having the network predict the concentration parameters of a Dirichlet distribution instead of categorical probabilities. This allows the model to learn uncertainty in a single forward pass, eliminating the need for sampling-based methods such as Monte Carlo dropout while maintaining calibration quality. Conformal prediction provides a distribution-free framework for generating prediction sets with guaranteed coverage [330]. In medical image segmentation, conformal methods have been adapted to produce voxel-wise or region-wise prediction sets that ensure the true label is contained within the predicted set at a predefined confidence level (e.g., 95%). Unlike other calibration approaches, conformal prediction directly controls the false prediction rate, making it particularly attractive for assisting clinical decision-making.

Despite these advances, challenges remain. Probabilistic methods often incur higher computational costs, particularly for Bayesian inference or ensemble-based approaches, which may limit real-time clinical applicability. Moreover, calibrating uncertainty estimates to ensure reliability across diverse datasets and modalities remains an open problem.

## 5.6 From 2D to 3D and 4D Segmentation

While early segmentation models primarily targeted 2D image slices, there is growing interest in extending these approaches to 3D volumetric and 4D dynamic segmentation to better capture anatomical context and temporal coherence. Advances in computational hardware, particularly increased GPU memory, have facilitated notable progress in this area. Existing methods of 3D segmentation generally fall into two categories: (1) extending 2D architectures into 3D using auxiliary strategies or modules, such as VNet, Swin UNETR, and UNETR++ [331]; and (2) treating 3D volumes as sequential 2D data—akin to video—to model inter-slice relationships, exemplified by models such as SAM2 [35], MedSAM2 [332] and MedSAM-2 [333]. Beyond these, streaming-memory schemes (e.g., SAM-2's key–value caches) propagate masks across frames with minimal re-prompting and have begun to be applied to cine ultrasound/MRI and slice-wise 3D propagation [45, 321]. Despite these developments, key challenges remain: (1) effectively managing the substantial computational and memory demands of high-resolution volumetric data; (2) addressing overfitting risks due to the limited availability of annotated 3D medical datasets; and (3) enhancing cross-modality and cross-institutional generalization to ensure robustness and clinical applicability across diverse imaging protocols and patient populations.

In 4D medical image segmentation, a central challenge lies in effectively modeling the complex spatiotemporal dependencies inherent in volumetric time-series data. Existing approaches have adopted a range of strategies to address this issue. These include 4D convolutions for simultaneous spatial-temporal representation learning [334], recurrent architectures such as long short-term memory (LSTM) networks to model temporal dynamics [335], and transformer-based frameworks that enable joint attention across spatial and temporal dimensions [336]. More recently, structured state space models, particularly those based on Mamba blocks [337] have demonstrated potential for efficient sequence modeling in high-dimensional medical data [338]. Despite these developments, several key challenges persist in 4D medical image segmentation. First, scalability and efficiency remain critical concerns, as many existing approaches—particularly those utilizing 4D convolutions or attention-based architectures—incur substantial computational and memory costs, limiting their feasibility for large-scale clinical deployment or real-time applications. Second, ensuring temporal consistency is nontrivial: minor spatial segmentation errors can accumulate over time, leading to temporal drift and compromising the reliability of longitudinal assessments. Third, generalization and robustness continue to pose difficulties, as current models often struggle to adapt across varying patient populations, imaging protocols, and modalities due to limited annotated datasets and the inherent heterogeneity of 4D medical images. Lastly, the development of high-quality, annotated dynamic imaging datasets remains an important objective. Although several studies have constructed 4D spatiotemporal datasets [339-341], there is still a notable lack of large-scale 4D datasets specifically designed for medical image segmentation.

Addressing these challenges necessitates the design of novel architectures that can more effectively capture spatiotemporal dependencies while enhancing computational efficiency and model robustness through the full exploitation of dataset-derived prior knowledge. Promising future directions include hybrid frameworks that integrate the structured efficiency of state-space models with the representational capacity of attention mechanisms, as well as the adoption of self-supervised and cross-modal learning strategies to mitigate data scarcity and domain shift. Further enhancements may be realized by coupling streaming memory mechanisms with deformable registration and uncertainty-aware re-initialization strategies, as well as by adopting temporally consistent evaluation metrics to ensure stable performance over long sequences.

## 5.7 From model-invocation to segmentation agents

Medical segmentation is mostly performed as one-shot inference, where a model produces a mask without explicit reasoning or interaction. However, to better capture clinical intent and ensure reliable outcomes, segmentation can be supported by agents that can plan tasks, compose tools, interact with clinicians, and audit results before downstream use. Language-driven interactive systems (for example, LIMIS [342]) adapt grounded promptable segmentation models so experts can revise masks via text instructions rather than pixel-level edits, enabling iterative correction under distribution shift and reducing manual effort. Building on this, multi-modal medical agents such as MMedAgent [343] instruction-tune a large language model to select and parameterize segmentation tools on demand (for instance, invoking MedSAM for organ or lesion masks), thereby closing the loop among goal understanding, tool invocation, and report synthesis.

At the vision-backbone layer, foundation segmentation model can provide the agent's operational substrate. MedSAM offers broad 2D coverage across modalities and anatomies for promptable segmentation, while SAM-Med3D [344] extends promptability to volumetric data with large-scale 3D pretraining; both function as reusable tools within agent pipelines. For open-vocabulary localization prior to fine segmentation, agents can chain Grounding DINO [345] to locate text-specified targets and then call SAM to delineate precise masks; recent Grounded-SAM [346] variants integrate tracking and streaming memory (via SAM-2) to stabilize 4D delineation with minimal re-prompting, which is practical for ultrasound or cine MRI.

A complementary line treats the agent as a quality gate. The Hierarchical Clinical Reasoner (HCR) [347] steers an LLM through staged clinical reasoning, including knowledge recall, visual/shape inspection, anatomical plausibility, and decision, to judge clinical acceptability of AI masks beyond pixel overlap, surfacing boundary or pathology errors that Dice may miss and producing explanatory rationales for audit. Looking ahead, segmentation agents should maintain prompt protocols (task–modality–region templates with automatic re-prompt on low confidence), exploit foundation tools with parameter-efficient tuning for site-specific adaptation, fuse uncertainty maps from the segmentation model to trigger re-segmentation or human review, and log tool chains and rationales for regulatory traceability. Together, these advances position agents as clinician-in-the-loop orchestrators that transform segmentation from a single prediction into a planned, checked, and dialog-refinable process.

Beyond the aforementioned topics, it is important to remember that the ultimate goal of medical image segmentation algorithms is to minimize human time and effort in clinical workflows. Consequently, evaluating segmentation performance should not be limited to conventional accuracy metrics but should also consider practical, time-related factors. Several key points merit attention:

(i) The most meaningful measure of a segmentation model's effectiveness in real-world clinical applications is the total human time required per study. For fully automated methods, this corresponds to the time needed to review and correct the segmentation until it is clinically acceptable. For interactive methods such as MedSAM, this includes both the time spent providing prompts and the time required for any post hoc corrections. Notably, very few studies have quantified this time-per-study component for any model, particularly for foundation models.

(ii) Such assessments must be performed separately for each new application; results from one or two use cases are insufficient to support broad generalizations. Commonly used evaluation metrics—such as Dice coefficient, Hausdorff Distance, False Positive/Negative rates, Precision, and Recall—rely on the availability of ground truth annotations and cannot directly quantify human time requirements for several reasons [348]. Moreover, these metrics have a non-linear relationship with clinical acceptability. For example, a Dice score of 0.95 generally denotes excellent quality for large, well-defined objects, whereas 0.85 would indicate inferior quality. In contrast, for small, slender, or spatially sparse structures such as the esophagus, a Dice of 0.85 may still represent excellent performance [2, 348, 349].

(iii) At present, no standardized tools exist to estimate this human time requirement without actually performing the manual corrections of auto-segmentations or executing interactive segmentation for every case using large models [350].

## 6. Conclusion

In this survey, we revisited the fundamental question: "*Is the medical image segmentation problem solved?*" —a question posed in the context of the rapid advances driven by deep learning. To address this, we first reviewed representative works from past decades, tracing the key milestones in the evolution of this field and the collective pursuit of advancing segmentation accuracy, efficiency, and applicability. Our conclusion is that, while significant progress has been made and we are closer than ever to this goal, the "holy grail" of fully solved medical image segmentation remains elusive. We therefore identified and discussed seven key challenges that will shape the trajectory of future research.

In summary, our findings suggest that future progress will hinge on two major directions. First, there is a pressing need to develop scalable and generalizable models, particularly by integrating large vision–language models capable of handling multi-modal inputs, varying dimensionalities (2D, 3D, 4D), and diverse segmentation targets, including organs, lesions, and other anatomies of clinical relevance. Second, achieving efficiency in both training and deployment remains critical—this entails reducing dependence on labeled data, minimizing computational overhead, enabling real-time inference on resource-constrained hardware, and ensuring high segmentation accuracy with interpretable outputs that can directly support clinical decision-making. We hope this survey provides valuable insights and inspires continued innovation in the quest for truly robust and universally applicable medical image segmentation solutions.

## Acknowledgements

The study was supported by the US National Institutes of Health (R01 CA240808, R01 CA258987, R01 EB034691, and R01 CA280135). The author, Guoping Xu, would also like to acknowledge the valuable discussions and assistance of Hua-Chieh Shao and Yubing Tong.

learning of invariant feature hierarchies with applications to object recognition' (IEEE, 2007, edn.), pp. 1-8